\documentclass[%
 reprint,
 amsmath,amssymb,
 aps,
floatfix
]{revtex4-2}
\usepackage[utf8]{inputenc}
\usepackage{graphicx} 
\usepackage{subcaption}
\usepackage{dcolumn} 
\usepackage{bm}
\usepackage{dcolumn}
\usepackage{color}
\usepackage{aas_macros}
\usepackage{multirow}
\usepackage{changepage}
\usepackage{hyperref}

\definecolor{mypurple}{rgb}{0.5, 0, 0.85}
\definecolor{Hbetta}{rgb}{0,0.92,1}
\definecolor{myblue}{rgb}{0, 0.2, 0.85}
\definecolor{cadmiumgreen}{rgb}{0.0, 0.42, 0.24}

\definecolor{review}{rgb}{1.0, 0, 0.0}
\definecolor{close}{rgb}{1.0, 0.5, 0.0}
\definecolor{final}{rgb}{.1, 0, 1}

\definecolor{mynote}{rgb}{0.6, .3, 0}

\newcommand{\eq}[1]{Eq.~(\ref{#1})}
\newcommand{\fig}[1]{Fig.~\ref{#1}}

\begin{document}

\title{Modeling Core-Collapse Supernovae Gravitational-Wave Memory in Laser Interferometric Data}

\author{Colter~J.~Richardson} \author{Michele~Zanolin} \affiliation{
  Embry-Riddle Aeronautical University, \\ 3700 Willow Creek Rd., Prescott, Arizona 86301 }

\author{Haakon~Andresen} \affiliation{ Max Planck Institute for Gravitational Physics (Albert
  Einstein Institute), \\ Potsdam Science Park Am Mühlenberg 1 D-14476 Potsdam }

\author{Marek~J.~Szczepa\'nczyk} \affiliation{ University of Florida, Department of Physics, \\ 2001
  Museum Road Gainesville, FL 32611 }

\author{Kiranjyot~Gill} \affiliation{ Harvard University, Department of Astronomy, \\ Center for
  Astrophysics, 60 Garden Street, MS-10 Cambridge, MA 02138 }

\author{Annop~Wongwathanarat} \affiliation{ Max Planck Institute for Astrophysics,
  \\ Karl-Schwarzschild-Str. 1, D-85748 Garching }

\date{\today}

\begin{abstract}
We study the properties of the gravitational wave (GW) emission between $10^{-5}$ Hz and $50$ Hz (which we refer to as low-frequency emission) from core-collapse supernovae, in the context of studying such signals in laser interferometric data as well as performing multi-messenger astronomy. We pay particular attention to the GW linear memory, which is when the signal amplitude does not return to zero after the GW burst. Based on the long-term simulation of a core-collapse supernova of a solar-metallicity star with a zero-age main sequence mass of 15 solar masses, we discuss the spectral properties, the memory's dependence on observer position and the polarization of low-frequency GWs from non (or slowly) rotating core-collapse supernovae. We make recommendations on the angular spacing of the orientations needed to properly produce results that are averaged over multiple observer locations by investigating the angular dependence of the GW emission. We propose semi-analytical models that quantify the relationship between the bulk motion of the supernova shockwave and the GW memory amplitude. We discuss how to extend neutrino generated GW signals from numerical simulations that were terminated before the neutrino emission has subsided. We discuss how the premature halt of simulations, and the non-zero amplitude of the GW memory can induce artefacts during the data analysis process. Lastly, we also investigate potential solutions and issues in the use of taperings for both ground and space-based interferometers.
\end{abstract}

\maketitle

\section{Introduction} \label{sec:Introduction}
Recent theoretical predictions of GWs emitted by core-collapse supernovae \citep{2012A&A...537A..63M, 2015PhRvD..92h4040Y, 2016ApJ...829L..14K, 2017MNRAS.468.2032A, 2018ApJ...865...81O, 2019MNRAS.486.2238A, 2019MNRAS.487.1178P, 2019ApJ...876L...9R, 2019MNRAS.489.2227V, 2020MNRAS.494.4665P, 2020PhRvD.102b3027M, 2021MNRAS.503.3552A} typically focus on frequencies above $\approx 10$ Hz in the time between core-bounce and the first second(s) after shock revival, referred to as the post-bounce phase. The turbulent mass-motions behind the stalled supernova shock predominantly leads to GW emission in a frequency range of tens of Hz to the order of kHz during the post-bounce phase. However, lower frequency GWs are expected from asymmetric emission of neutrinos \citep{1978ApJ...223.1037E, 1996PhRvL..76..352B, 1997A&A...317..140M, 2007ApJ...655..406K, 2009ApJ...697L.133K, 2009ApJ...704..951K, 2011ApJ...736..124K}, and the aspherical ejection of matter \cite{2009ApJ...694..664M, 2009ApJ...707.1173M, 2010CQGra..27s4005Y}. These two processes can produce GWs peaked at frequencies around and below a few Hz \citep{2009ApJ...694..664M, 2009ApJ...707.1173M, 2013ApJ...766...43M}.

The first evidence for GW memory from core-collapse supernovae emerged in two-dimensional numerical simulations \citep{2009ApJ...707.1173M, 2013ApJ...766...43M, 2015PhRvD..92h4040Y}. However, it was understood that the symmetry constraints of such simulations made them prone to highly aspherical explosions. Indeed, the relatively large GW memory seen in two-dimensional simulations so far has not been observed to the same degree in recent three-dimensional simulations of slowly rotating progenitors \citep{2012A&A...537A..63M, 2019MNRAS.487.1178P, 2020PhRvD.102b3027M}. The axial symmetry of rapidly rotating progenitors also has prominence for axisymmetric explosions and large memory amplitudes, but we leave these rarer progenitors for future investigations. Independent of the progenitor, the exact characteristics, and asymptotic values of the GW emission of a core-collapse supernova can only be determined from three-dimensional numerical simulations. For this reason, the illustrative numerical simulation we use here is three-dimensional.

The amplitude of the low-frequency emission can be one to two orders of magnitude larger than the emission caused by the turbulence behind the shock \cite{2020MNRAS.492.4613O}. It is worth noting, however, that the energy contained in the low-frequency component is expected to be much less, because the energy density in the frequency domain is proportional to the square of the frequency. In the GW literature two types of memory have been identified: non-linear \cite{2020PhRvD.101b3011H} and linear \citep{Maggiore:2007ulw}, but only the second type is expected to be relevant for core-collapse supernovae.

In this work, we study the properties of the GW memory using as an example \citep{2015A&A...577A..48W}, a $15 M_{\odot}$ core-collapse supernova simulation with no rotation. \citep{2015A&A...577A..48W} simulated the evolution of a 3-D explosion, from core-bounce to shock break out, but with approximate neutrino physics. Furthermore, the neutrino transport was switched off after 1.3 s. While the simulation is approximate, it provides insight into the propagation of the blast-wave and enables us to directly calculate the memory associated with the ejection of stellar material. We extend the neutrino generated signal beyond 1.3 seconds, with an analytical extension, which will allow us to parametrize some of the uncertainty of the problem and explore different possible continuations of the signal.

The laser interferometer GW data analysis community is developing data analysis strategies for core-collapse supernovae (see for example \citep{2021arXiv210406462S, 2020PhRvD.101h4002A, 2021PhRvD.103f3006B, 2018PhRvD..98l2002A, 2019PhRvD..99f3018R}). These efforts have focused, so far, on the frequency range where ground-based detectors are most sensitive rather than the band where we expect the effects of the linear memory to be important. However, current and future instruments are expected to have better sensitivities in the frequency range where the memory is the dominant component (see \ref{sec:signal}).  The goal of this paper is to study the properties of the memory focusing on the scientific goals achievable with laser interferometers. In order of decreasing frequencies ranges, the relevant instruments are Advanced LIGO \citep{2018LRR....21....3A}, Cosmic Explorer \cite{P2100003}, Einstein Telescope \cite{2011CQGra..28i4013H}, TianGO \cite{2020PhRvD.102d3001K}, DECIGO \cite{2017JPhCS.840a2010S}, and LISA \cite{2021arXiv210801167B}.

Recently, the memory specifically associated with asymmetric neutrino emission was studied in \citep{2020ApJ...901..108V}. They find that the neutrino generated GW signal from a galactic supernova will be above the noise floor of AdvancedLIGO, the Einstein Telescope, and proposed space-based detectors. However, this conclusion is for the total signal and not the memory specifically. It was also not discussed whether or not the sudden termination of the GW signal at a non-zero value introduced artifacts in the Fourier analysis or how their result depends on the source orientation.

The development of data analysis algorithms in preparation for extraordinary events like the next galactic SN is based on injecting simulated GWs into the data stream. For the results to be accurate, the waveforms should not be distorted, and results are often presented as averaged over different source orientations, with the caveat that the number of orientations should sufficiently represent the signal variability. For the first issue, a source of concern is the temporal truncation of the GW when the simulation is ended. If the signal is injected as is in the data, it produces non-physical broadband distortions to the reconstructed spectrum. We discuss extensions of the GW signal past the truncation of the simulation by the addition of a tapering function that gradually brings the signal to zero. Sometimes the development of the memory is not completed by the time the truncation is performed, as indicated by a non-zero slope at the end of the simulation, so we also investigate the impact of analytical continuations to an asymptotic value, for the sensitivity band of different, existing and proposed, interferometers.

\begin{figure*}
    \includegraphics[width=0.95\textwidth]{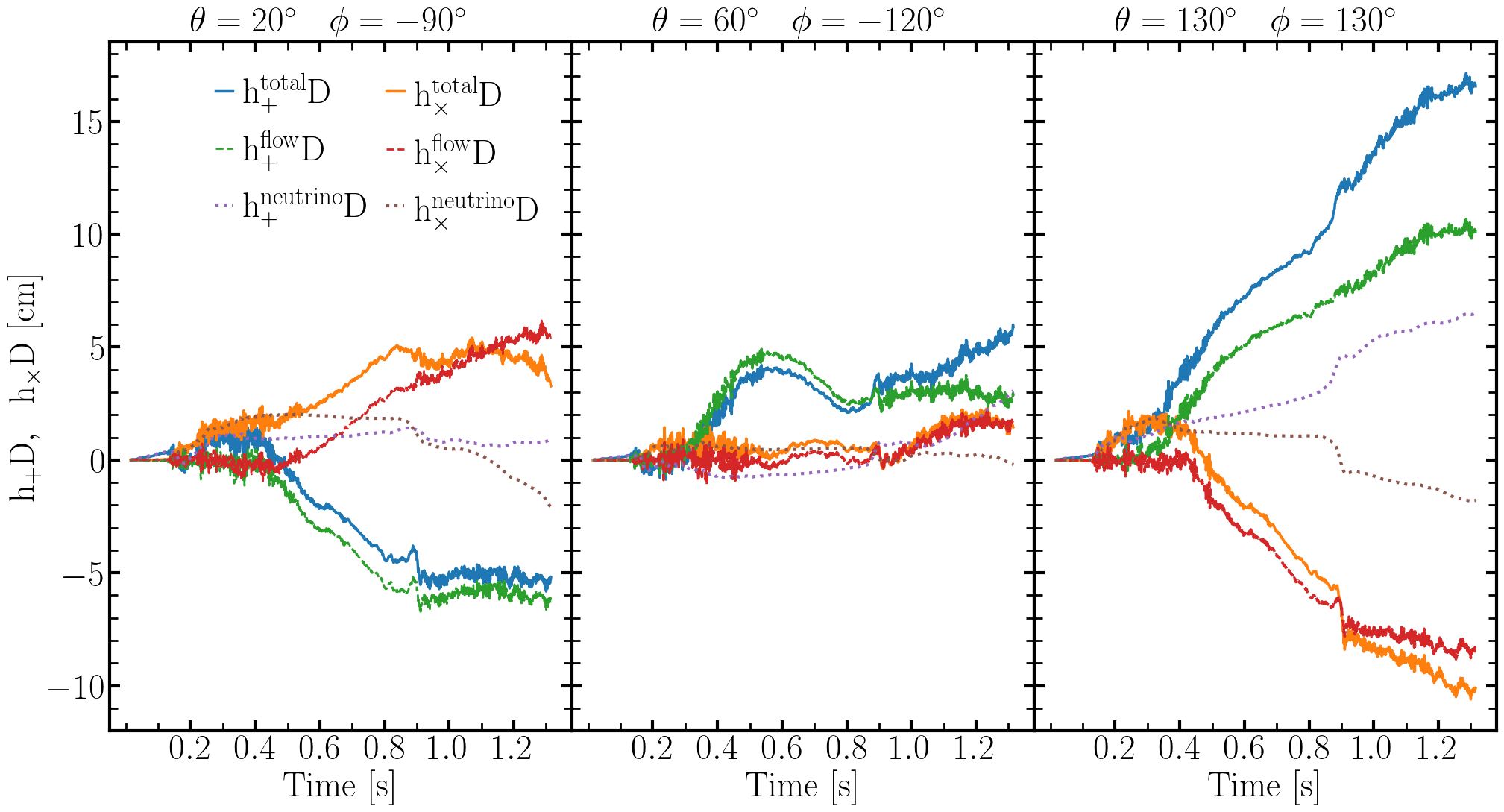}
    \caption{GW strain multiplied by the distance to the source for three different observer orientations, relative to the spherical coordinate system of the hydrodynamic simulation. Each column represents one observer direction, which is denoted by the azimuthal ($\phi$) and polar ($\theta$) angle above the column. Time is given in seconds after core bounce.}
    \label{fig:amps}
\end{figure*}

In the context of the GW signal contributed by the neutrino signal, more extensions have also been investigated in \citep{2021arXiv210505862M}. The extensions in \citep{2021arXiv210505862M} are described by an analytical model dependent on the time-evolution of the anisotropy of the neutrino emission, which is important in the production and angular dependence of the GW signal. Our work instead investigates a physically motivated constant anisotropy parameter discussed in more detail in section \ref{sec:Extension}.

In this paper we discuss polarization properties for the total memory produced by neutrinos and hadronic matter, we discuss the angular dependence of the total memory and provide interpretations of the hadronic matter GW memory production with toy models for the matter ejection. We also discuss how the relevant band for the memory not only depends on the memory's central frequency but also the noise floor as well as the frequency content of other GW production mechanisms in CCSNe.

In section \ref{sec:Numerical} we discuss the simulation which our analysis is based on. In section \ref{sec:CCSN} we discuss the processes that lead to GW emission and make the distinction between linear and non-linear memory. In section \ref{sec:signal} we discuss the frequency range where the memory is a dominant contribution (roughly up to $50$ Hz) In section \ref{sec:Angular Dependence Interpretation} the angular dependencies are studied and compared to three toy model explosions and recommendations are provided for the number of source orientations to use in data analysis studies. In section \ref{sec:low} we discuss the polarization of the GW signal in the low-frequency limit. In sections \ref{sec:Extension} and \ref{sec:Tapering} we discuss the impact of different extensions and tapering assumptions, both based on signal processing consideration and physical considerations, including the scenario where the memory has not fully developed yet for various noise spectral densities.

\section{A Numerical Model} \label{sec:Numerical}
To study the low-frequency GW emission from core-collapse supernova, we investigate the emission from model W15-2 presented in \citep{2010A&A...514A..48W, 2013A&A...552A.126W, 2015A&A...577A..48W}. Model W15-2 is a fully three-dimensional numerical simulation of the evolution of a core-collapse supernova from 15 ms after bounce until the shock breaks out of the stellar surface. The simulation was carried out in two phases, which we will refer to as the first and second phases. In short, the first phase lasts until $1.3$ s after core bounce and during this phase the shock is successfully revived. After $1.3$ s, the data from the first phase was mapped onto a new grid which made it possible to follow the shock as it propagated through the stellar progenitor. Due to computational limitations, the second phase did not include neutrino transport (unlike the first phase). The simulation was carried out with the explicit finite volume code \textsc{Prometheus} \citep{1991ApJ...367..619F, 1991ESOC...37...99M, 1991A&A...251..505M} using the axis free Yin-Yang grid \citep{2004GGG.....5.9005K, 2015A&A...577A..48W}. Model W15-2 is based on the non-rotating model s15s7b2 of \citep{1995ApJS..101..181W} which had a ZAMS mass of 15 solar masses.

During the first phase, the two grid patches of the Yin-Yang grid had $400$ cells in the radial direction, $74$ in the polar direction and $137$ in the azimuthal direction, which corresponds to an angular resolution of $2$ degrees. The outer radial boundary was placed at $18,000$ km. The high-density central region of the PNS was excised from the computational domain and replaced by an inner radial boundary condition and a point-like mass at the origin. The radial position of the inner boundary decreased with time from an initial value of 65 km to a radius of $15$ km at the end of the first simulation phase. Moving the inner boundary further and further inwards mimics the contraction of the PNS (see \citep{2008A&A...477..931S, 2010A&A...514A..48W} for details about the boundary prescription).

Neutrino radiation transport was performed with an energy integrated Ray-By-Ray transport scheme \cite{2006A&A...457..963S}. The Ray-By-Ray approximation is known to result in stronger local variation in the neutrino heating rate \citep{sumiyoshi_15,skinner_16,just_18,glas_19}. Small scale and local artefacts are averaged out when performing the volume integrals necessary to calculate the GW signal and are, therefore, not too troublesome for our work. In a similar way, the angular grid resolution constrains how finally we can resolve the neutrino signal. Again, since we are interested in the spatially integrated and slowly evolving quantities, small and short variations are not a source of concern for our study. The reader should keep in mind that the simulation uses an approximate neutrino scheme and that the signal presented must be interpreted with some care. The overall, global properties are most likely well represented, but the exact details are uncertain. In the future, simulations with more sophisticated neutrino transport will shed light on this very issue. This is one of the reasons why our analysis focuses on the overall properties of these kinds of signals and not the small details of the particular example signal we have chosen. 

The second phase retained the angular grid from the first phase, but the radial grid was refined. The outer radial boundary was moved to $3.3 \times 10^8$ km. The inner radial boundary was set to 500 km and was moved outwards as the shock propagated further and further from the center of the computational domain (see \cite{2015A&A...577A..48W} for details). Importantly, neutrino transport was not included in the second phase.

\section{GW Emission}
In this section, we give an overview of the GW emission from a slowly rotating core-collapse supernova and discuss the properties of the low-frequency emission and GW memory of model W15-2.

\subsection{Frequency Content} \label{sec:CCSN}
Since the advent of fully three-dimensional numerical simulations with sophisticated neutrino transport, predictions for the GW signals emitted by core-collapse supernova (for slowly/non rotating progenitors) have started to converge. It has been established that the hydrodynamic instabilities operating in the post shock region and within the PNS are the main sources of GW emission above a few tens of hertz. The neutrino driven convection manifests as high entropy bubbles rising from the PNS to the shock front, they arise due to the neutrino heating at the bottom of the post shock layer. The turbulent fluid flow can excite oscillation modes in the PNS, which in turn emit GWs. The stalled accretion shock instability (SASI) develops through an advective-acoustic cycle, entropy and vorticity perturbations from the shock are advected through the post shock layer to the PNS. Additionally, the SASI coherently modulates the flow on large scales which leads to GW emission. Oscillations of the PNS typically leads to emission above 300 Hz \citep{2009ApJ...707.1173M, 2009ApJ...694..664M, 2013ApJ...766...43M} and emission between 50 and 250 Hz \citep{Andresen:2015gyi, 2016ApJ...829L..14K, 2017MNRAS.468.2032A} has been associated with the SASI. The expected central frequency of the two signal components has been estimated, either by physical arguments \citep{2009ApJ...707.1173M, 2009ApJ...694..664M, 2013ApJ...766...43M, 2019MNRAS.486.2238A} or by mode analysis \citep{2017PhRvD..96f3005S, 2018MNRAS.474.5272T, 2018ApJ...861...10M, 2019PhRvL.123e1102T, 2019MNRAS.482.3967T, 2020PhRvD.102b3028S}. While a discussion around the details is still ongoing, the different approaches find similar results. According to the fitting formulas derived in \citep{2019PhRvL.123e1102T}, the central frequency of the emission from the PNS can be expressed as

\begin{align} \label{eq:fgwpns}
    f_{GW}^{PNS} = & 12.4 \times 10^{5} Hz \bigg(\frac{m_{PNS}}{r_{PNS}^{2}}\bigg) - 378 \times 10^{6} Hz \bigg(\frac{m_{PNS}}{r_{PNS}^{2}} \bigg)^2 \nonumber \\
    & + 4.24 \times 10^{10} Hz \bigg(\frac{m_{PNS}}{r_{PNS}^{2}} \bigg)^3,
\end{align}
and the typical frequency of the emission associated with SASI activity is given by
\begin{equation} \label{eq:fgwsasi}
    f_{GW}^{SASI} = 2 \times 10^{5} Hz \sqrt{\frac{m_{sh}}{r_{sh}^3}} - 8.5 \times 10^{6} Hz \bigg(\frac{m_{sh}}{r_{sh}^3}\bigg).
\end{equation}

Here $m_{PNS} = M_{PNS}/M_{\odot}$, $r_{PNS} = R_{PNS}/1\mathrm{km}$, $m_{sh} = M_{sh}/M_{\odot}$, and $r_{sh} = R_{sh}/1\mathrm{km}$, where $M_{PNS}$, $R_{PNS}$, $M_{sh}$, and $R_{sh}$ are the mass of the PNS, the radius of the PNS, total mass behind the shock, and the average shock radius, respectively. The total mass behind the shock can be approximated by the PNS mass because most of the mass is confined within the central object. The PNS tends to have a radius of $20-60$ km and a mass of approximately $1-2$ solar masses. Before shock revival, the average shock radius will typically be $\approx 100-200$ km. If the PNS mass is $1.4$ solar masses, the PNS radius $60$ km and the shock radius $150$ km, then the $f_{GW}^{SASI} \approx 125$ Hz and $f_{GW}^{PNS} \approx 430$ Hz (Note that the latter will evolve as the PNS cools and contracts, it approximately increases linearly with time and can reach values of $\approx 2$ kHz).

\begin{figure*}
    \centering \includegraphics[width=0.95\textwidth]{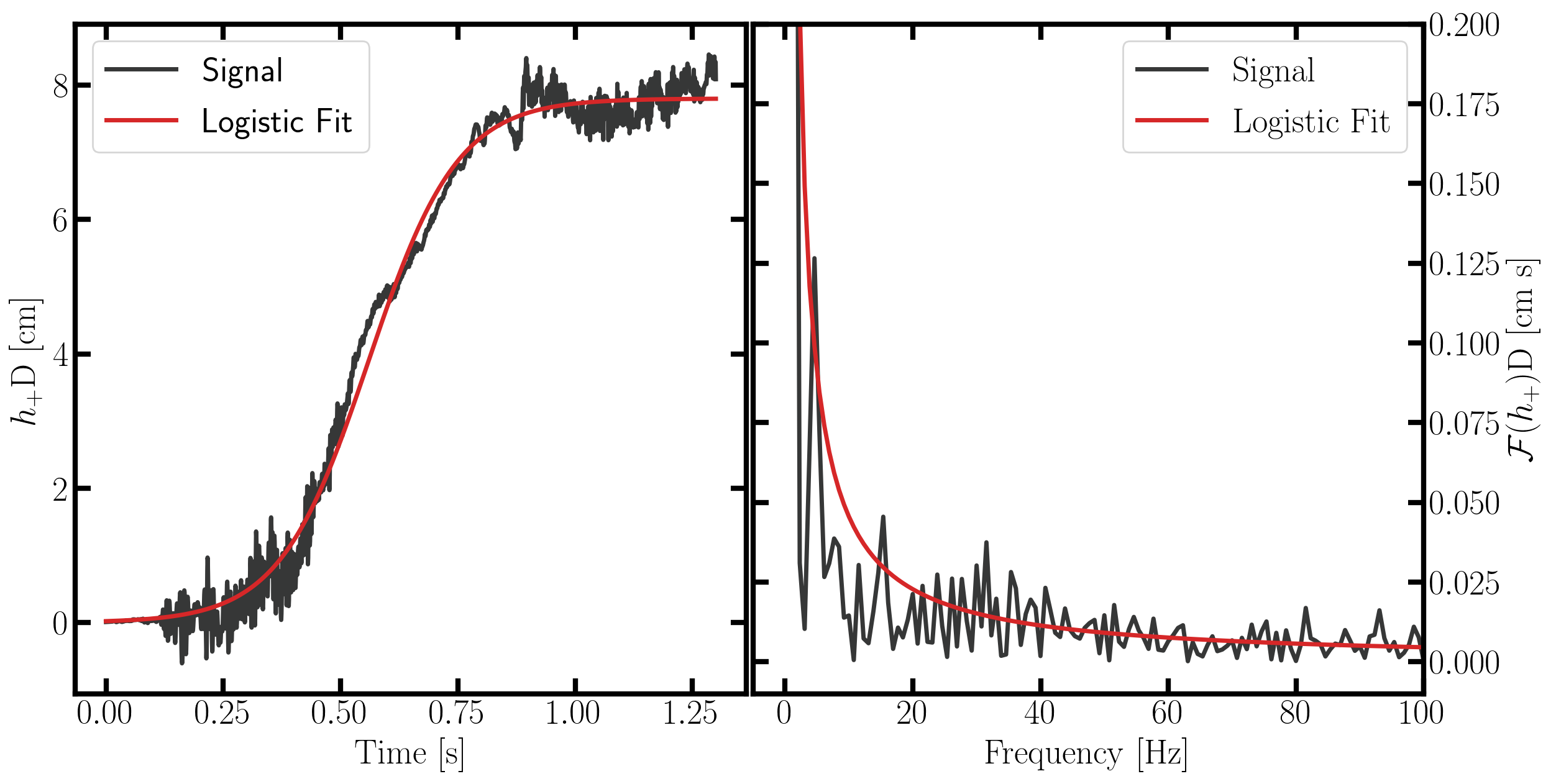}
    \caption{Left panel: $h_+ D$ for an arbitrary observer (in black) and the best fit of the signal with a logistic function (in red), see section~\ref{sec:signal}.  Left panel: The Fourier transform of the curves in the left panel, with the same color coding as the right-left panel. The good agreement between the red and black curve, below $\approx 20$ Hz, shows that the low-frequency part of the signal is dominated by the secular ramp-up of the memory.}
    \label{fig:Fitted Logistic Function}
\end{figure*}

In addition to the radiation emitted between $50$ Hz and $2$ kHz, asymmetric emission of neutrinos and non-spherical shock expansion (after shock revival) can lead to a secular increase in the absolute value of the GW strain. The amplitude of this emission can be orders of magnitude larger than the emission generated by the turbulent mass flow but is typically emitted at frequencies peaked below a few Hz. The low-frequency emission does not necessarily settle down to zero after the neutrino emission has subsided and after the shock has broken out of the stellar surface

\begin{equation}
    \label{eq:memdiff}
    h_{\times,+}(t_{i}) - h_{\times,+}(t_{f}) \ne 0,
\end{equation}

Here $t_i$ and $t_f$ denotes the time when the source starts and stops, respectively, emitting GWs. We do not discuss the non-linear memory, i.e the memory signals coupled with the GW bursts. While such effects are relevant for merging binary systems (see for example \cite{2018PhRvD..98f4031T}), core-collapse supernovae are not compact enough for this effect to be relevant \cite{1991PhRvD..44.2945W}.

\subsection{Low Frequency} \label{sec:signal}

In this section, we will give an overview of the GW emission properties below 50 Hz from model W15-2 (the emission at higher frequencies is not directly relevant for our work and has been described in detail by \citep{2012A&A...537A..63M}). We will use the terms {\it flow} and {\it neutrino} to separate the GWs generated by the fluid flow and the asymmetric emission of neutrinos, respectively.

\begin{figure*}
    \includegraphics[width=0.85\textwidth]{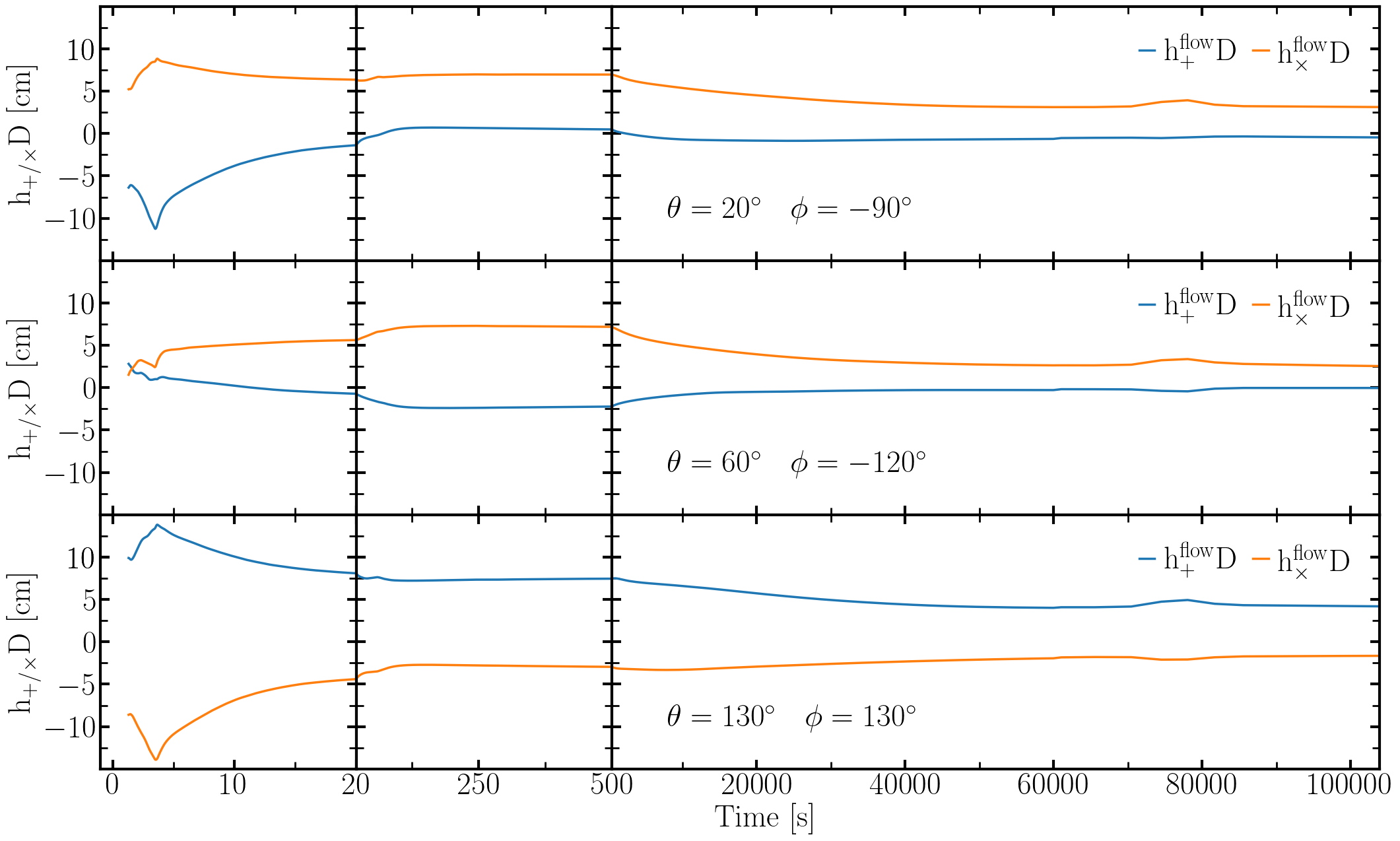}
    \caption{GW, for simulation phase two, strain multiplied by the distance to the source for three different observer orientations, relative to the spherical coordinate system of the hydrodynamic simulation. Each row represents one observer direction, which is denoted by the azimuthal ($\phi$) and polar ($\theta$) angle.  Time is given in seconds after core bounce.}
    \label{fig:amps_late}
\end{figure*}

The GW signal from the first simulation phase is shown in \fig{fig:amps}, the solid lines show the total signal and the dashed and dotted lines show the individual contribution from the fluid flow and the neutrino emission respectively. Each column in \fig{fig:amps} represents a different observer orientation. The observer position is given in terms of two angels, $\phi$ and $\theta$ which correspond to the azimuthal and polar angle, respectively, in the spherical coordinate system of the Yin-patch (see \cite{2015A&A...577A..48W} for a detailed description of the simulation grid) of the simulation grid. In general, the absolute value of signal amplitude lies between 5 to 15 cm and the amplitude of the low-frequency emission is typically larger than the amplitude of the high-frequency ($> 10$ Hz) signal, which can be seen as stochastic modulations on top of the secular time evolution of the signal. The observer situated at $(\theta, \phi) = (130^{\circ}, 130^{\circ})$ would observe a large positive $h_+$ and a large negative $h_\times$ (see the right column of \fig{fig:amps}). The situation would be reversed for an observer situated along the radial vector defined by $(\theta, \phi) = (20^{\circ}, 20^{\circ})$ and the total signal amplitude would be smaller for this observer (see the left column of \fig{fig:amps}).  The observer at $(\theta, \phi) = (60^{\circ}, 120^{\circ})$, see the middle column of \fig{fig:amps}, would observe the smallest signal out of the three observers and would at 1.3 s after bounce measure positive values for both polarization modes.

\fig{fig:Fitted Logistic Function} shows how the secular part of the signal dominates the emission between $10^{-5}$ and $\sim 50$ Hz. The left panel shows $h_+$ in the time domain for a randomly chosen observer and the best fit for a function of the form $f(t) = \frac{L}{1 + e^{-k (t - t_{0})}}$ to the signal (we refer to this as a logistic fit), where $L$ represents the final saturation value and $k$ represents the time scale. The Fourier transform ($\mathcal{F}$) of $h_+$ and the logistic fit is shown in the right panel of \fig{fig:Fitted Logistic Function}. The two curves are in good agreement below $\sim30$-$50$ Hz (the right panel of \fig{fig:Fitted Logistic Function}).

We show the GW signal from the fluid flow during the second simulation phase in \fig{fig:amps_late}, the signal is shown for the same three observer directions as in \fig{fig:amps}. Note that the x-axis of \fig{fig:amps_late} has been divided into three segments with different scales, this enables us to show the signals in their entirety without obscuring the changes occurring at relatively short time scales during the first $20$ s. The signal generated by the fluid flow undergoes rather large changes during the first $5$ s of the second simulation phase, often reaching a global maximum or minimum at around $5$ s after core bounce. The long-term evolution of the signal depends on the orientation of the observer, but in general the signal tends to reach a maximum value within the $500$ s after core bounce. After the signal reaches a global maximum, the signal typically tends towards zero or a small amplitude towards the end of the simulation.

If the source distance $R$ is large compared to the extent of the source (see \citep{2012A&A...537A..63M, Maggiore:2007ulw}), the total energy radiated by GWs through a sphere with radius $R$ is given by,
\begin{equation}
    E = \frac{c^{3}R^{2}}{16 G \pi} \int_{-\infty}^{\infty}\mathrm{d} t \int \mathrm{d} \Omega \big(\dot{h}_{\times}^{2} + \dot{h}_{+}^{2} \big),
\end{equation}

During the first simulation phase, model W15-2 radiates a total of $E \approx 1.97 \times 10^{43}$ erg as GWs between $25$ ms and $1.3$ s post bounce. Most of the energy is emitted in a frequency range of $100-2000$ Hz, because the energy scales as the frequency of the emission to the second power. The energy carried by the GW emission below $1$ Hz is $\approx 1.15 \times 10^{40}$ erg, which is three orders of magnitude smaller than the total radiated GW energy during this phase. During the second simulation phase a total of $\approx 3.01 \times 10^{39}$ erg are emitted as GWs generated by the fluid flow. It is important to point out, however, that interferometers are sensitive to the amplitude and not the energy, so the ratio of the signal amplitude to the amplitude of the noise is the most critical indicator for memory detectability (see the discussion of table \ref{tab:SNR}).

\subsection{Angular Dependence} \label{sec:Angular Dependence Interpretation}
\begin{figure*}
    \centering \includegraphics[width = 0.95\textwidth]{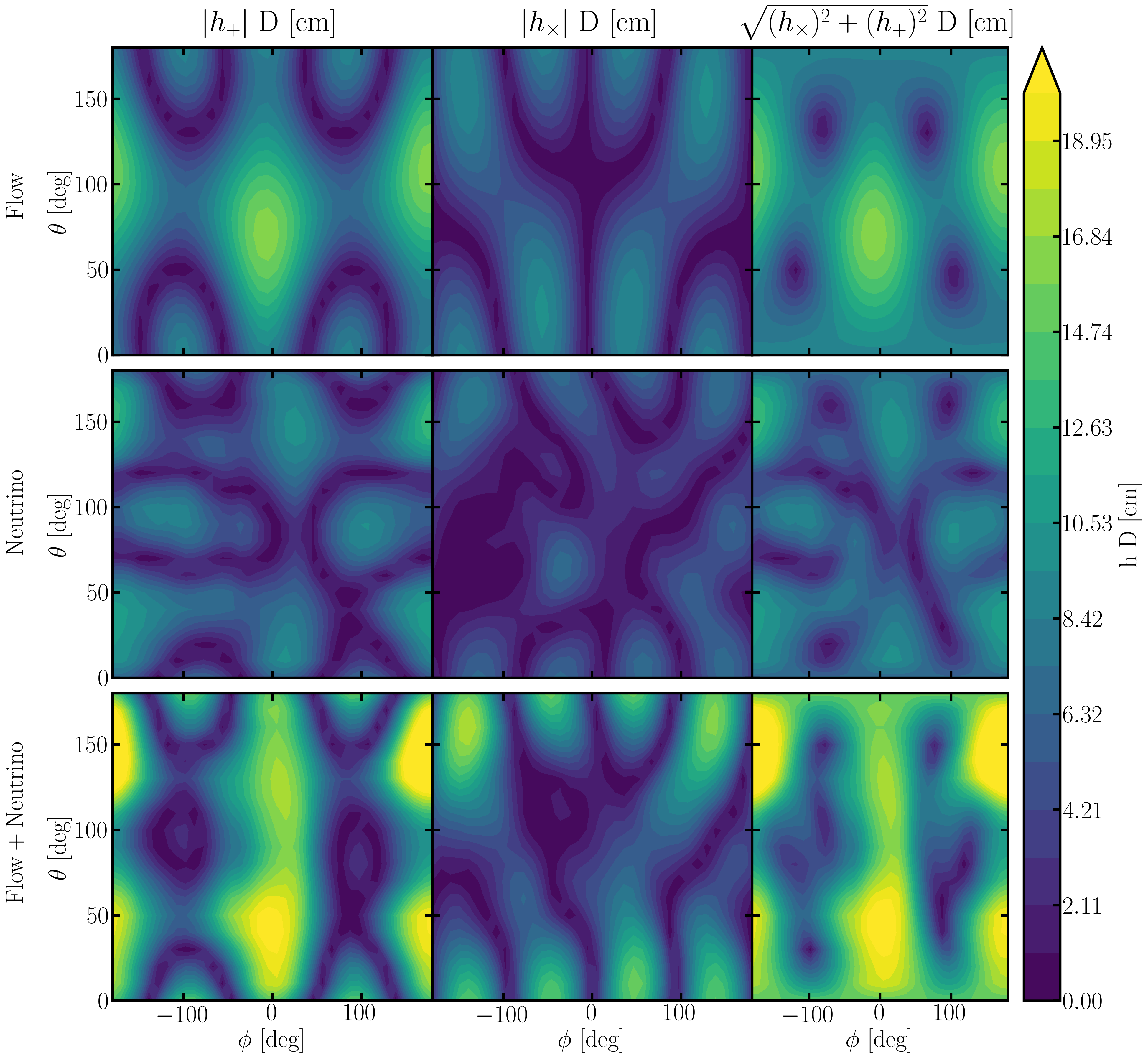}
    \caption{The angular dependency of the absolute value of the total GW signal from the simulation at 1.3 s after bounce. The left column shows the plus polarization mode, the middle column shows the cross polarization mode, and the right column shows the gauge invariant root sum square of the two
    modes. The top row shows the GWs generated by the fluid flow, the middle shows the GWs from asymmetric emission of neutrinos, and the bottom panel shows the total GW signal.}
    \label{fig:angular_dis}
\end{figure*}

\begin{figure}[h!]
    \centering
    \includegraphics[width = 0.40\textwidth]{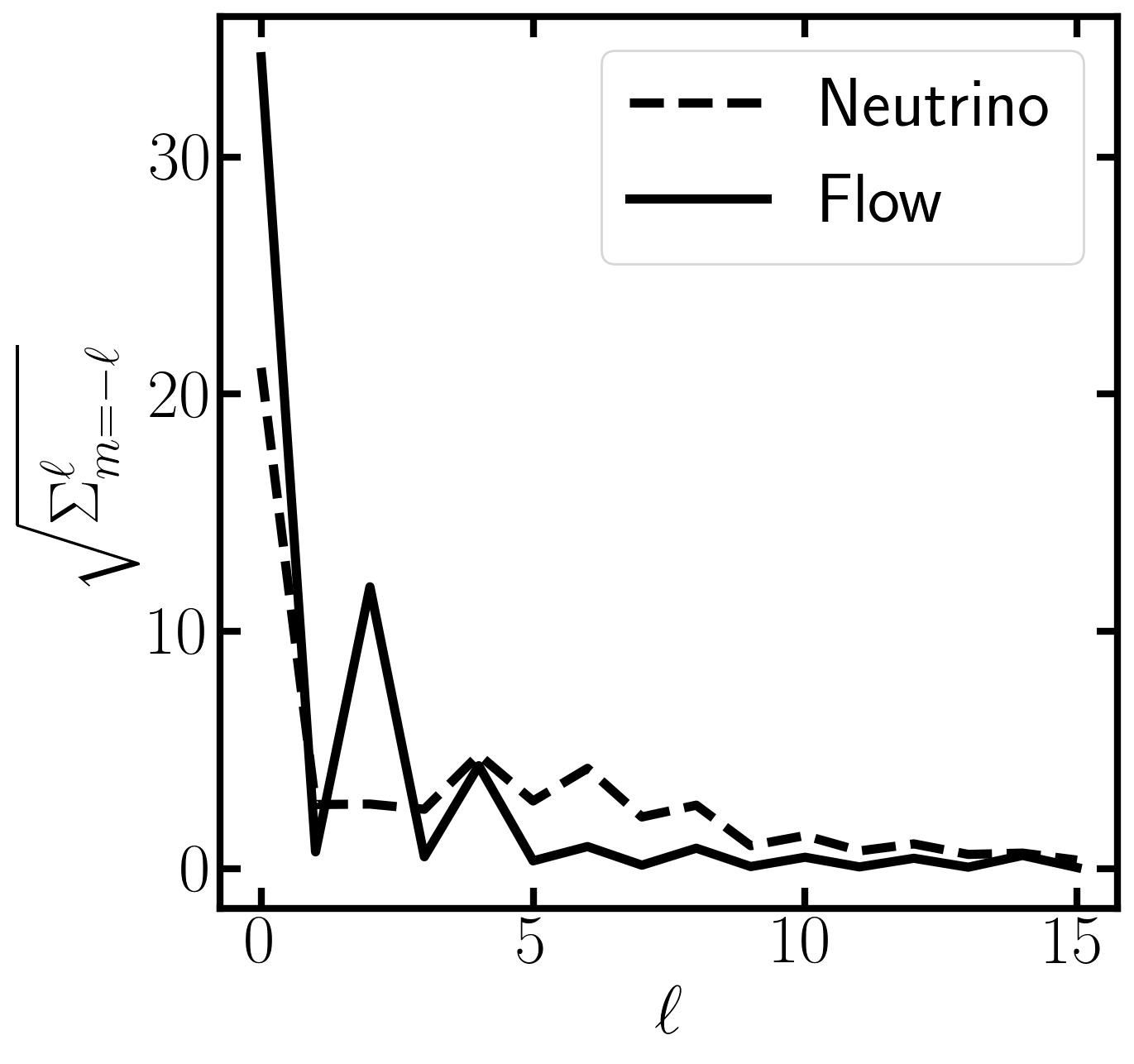}
    \caption{The spherical harmonic decomposition shown above illustrate how the flow has more energy located in the lower $\ell$ values than the neutrino..}
    \label{fig:SHE}
\end{figure}

\fig{fig:angular_dis} shows the angular dependence of the memory amplitude of the W15-2 simulation, at the end of the first simulation phase, for observer directions $\theta \in [0^{o}, 180^{o}]$ and $\phi \in [-180^{o},170^{o}]$. The signal generated by asymmetric neutrino emission shows a larger degree of variability than the GWs generated by the fluid flow. This is shown in \fig{fig:SHE}, where the spherical harmonic decomposition of the memory from the neutrino luminosity, which is less dominated by lower $\ell$, and the GWs from the flow, which is dominated by lower $\ell$, are shown for the first phase of the simulation. This indicates that there the neutrino memory is more variable. The coefficients of the spherical harmonic decomposition are defined as

\begin{equation}
    a_{\ell}^{m} = \frac{(-1)^{|m|}}{\sqrt{4 \pi (2 \ell +1)}} \int h_{+/\times}(\theta, \phi, t = 1.3 s) Y_{\ell}^{m}d\Omega
\end{equation}
where $h_{+/\times}(\theta, \phi, t = 1.3 s)$ is the amplitude of the memory at observer location $(\theta, \phi)$, and $Y_{\ell}^{m}$ is the spherical harmonic of degree $\ell$ and order $m$.

The ejection of material in model W15-2 generates a rather smooth and regular emission pattern that are dominated by a few large-scale structures. The angular dependence of the signal only varies over angular separations of several tens of degrees. This information is important when deciding how many orientations should be used to make statements about the average properties of the emission, which are customary in GW data analysis \citep{2020PhRvD.101h4002A}.

These maps also provide a total range of variability of the memory amplitude. For example, in the spectra plots presented in \fig{fig:No_LISA_1kpc_SNR} and \fig{fig:LISA_1kpc_SNR} we used the GW produced for the outgoing direction $\phi = -180$, and $\theta = 0$ where the amplitude of the memory is about 8 cm. This value is close to 7.15 cm, the average memory magnitude across all observer orientations, and is therefore a representation of the waveform family for the memory specifically. For an optimal orientation the signal spectra in \fig{fig:No_LISA_1kpc_SNR} and \fig{fig:LISA_1kpc_SNR}, below 30 Hz, would be about a factor 3 better.

To further interpret the angular dependence of the memory, we compare the GW emission of three toy models of the low-frequency emission of W15-2. The analytical models are constructed to mimic the properties of the matter ejected in W15-2, under the simplifying assumption of axis-symmetric matter ejection.

The first model we consider is \textit{Prolate} ejecta: two fluid parcels, each containing half of the asymmetric ejected mass (M), are ejected in the negative and positive z-direction (here the symmetry axis for the toy models), at distance $r=f(t)$ and with velocity $\dot{f(t)}$. At late times, assume that the velocity is approximately constant, in other words $\dot{f(t)} \approx v$, where $v$ is the velocity. The prolate model is defined by the density function
\begin{equation}
    \label{eq:Sym Pro}
    \rho = \frac{M}{2}\delta(z-f(t))\delta x \delta y + \frac{M}{2}\delta(z+f(t))\delta x \delta y.
\end{equation}

The second model we consider is the \textit{Oblate} model, which takes the form of an expanding ring in the plane perpendicular to the symmetry-axis. The matter ejection is restricted to radial motion in the $x$ - $y$ plane and has the density defined by
\begin{equation}
    \label{eq:Sym Ob}
    \rho = \frac{1}{2\pi} \frac{M}{r^2}\delta(\theta-\frac{\pi}{2}) \delta(r-f(t)).
\end{equation}

The third and last model we consider is the \textit{Spheroidal} explosion toy model, where the ejecta takes the form of an expanding ellipsoid. The matter ejection is restricted to a surface defined by $\frac{x^{2} + y^{2}}{a^{2}} + \frac{z^{2}}{c^{2}} = f^{2}(t)$, with $f(t)$ increasing over time. For spheroidal explosions, the mass radiates on a shell that resembles a prolate explosion, an oblate explosion, or a linear combination of both. The shape of the ellipsoid is controlled by the ratio $\frac{a(t)}{c(t)}$ (where both $a$ and $c$ are non-dimensional) and the density is given by
\begin{equation}
    \label{eq:Sym Sph}
    \rho = \frac{M}{2\pi a^{2}(t) c(t) f(t)} \delta \bigg(\frac{x^{2} + y^{2}}{a^{2}(t)} +
    \frac{z^{2}}{c^{2}(t)} - f^{2}(t)\bigg).
\end{equation}

To compute the GW emission from our analytical models, quadrupole moments based on the moment of inertia \citep{Maggiore:2007ulw} of each model, see appendix \ref{sec:ArbSph}, are computed for different observer directions. Note that the relationship between the moment of inertia of the three toy models is
\begin{equation}
    I_{Spheroid} = \frac{2}{3} \bigg(a^{2}(t) I_{Oblate} + c^{2}(t) I_{Prolate} \bigg),
\end{equation}
which we derive in appendix \ref{sec:ZSph}.

We perform a quantitative comparison of angular variability in the GW emission from our toy models with W15-2 by calculating the following reduced chi square ($\chi{2}$)
\begin{equation}
    \chi^{2}_{\nu} = \frac{\chi^{2}}{\nu} = \frac{1}{n - m} \sum_{i = 0}^{n}\frac{(E_{i} - O_{i})^{2}}{E_{i} + \delta},
\end{equation}
where $\nu = n - m$, $n$ is the number of angular locations we test and $m$ the number of estimated parameters. We vary three parameters, 1) a mass and velocity term ($M \times \alpha$, with $\alpha = \frac{G}{c^{4}}v^{2}$), 2) the explosion axis orientation ($\kappa$, azimuthal and $\mu$, polar), and in the case of the spheroidal explosion 3) the parameters $a$ and $c$ which determine the shape of the ellipsoid. $E_{i}$ is the expected signal (the amplitude of the memory in the simulation), $O_{i}$ is the signal from the toy models, and $\delta$ is a small number we added to remove numerical instabilities when $E_{i} = 0$ (we set $\delta = 0.1$).  Due to the addition of the $\delta$ in both the expected and observed signals, the $\chi_{\nu}^{2}$ values are slightly smaller than traditionally calculated $\chi_{\nu}^{2}$ values. The impact is, however, small: we calculated the same $\chi_{\nu}^{2}$ for the simulation against the same sheet with Gaussian noise ($\mu = 0$ and $\sigma = 2$) added both with and without the addition of $\delta$. With a Monte Carlo simulation we find that the $\langle\chi_{\nu}^{2}(\delta = 0)\rangle = 0.44$ with $\sigma = 0.05$ and $\langle\chi_{\nu}^{2}(\delta = 0.1)\rangle = 0.43$ with $\sigma = 0.05$. Here $\sigma$ indicates the $68\%$ confidence band assuming the gaussianity of the disturbances, which is not necessarily the case as discussed above.

\begin{figure*}
    \centering \includegraphics[width = 0.95\textwidth]{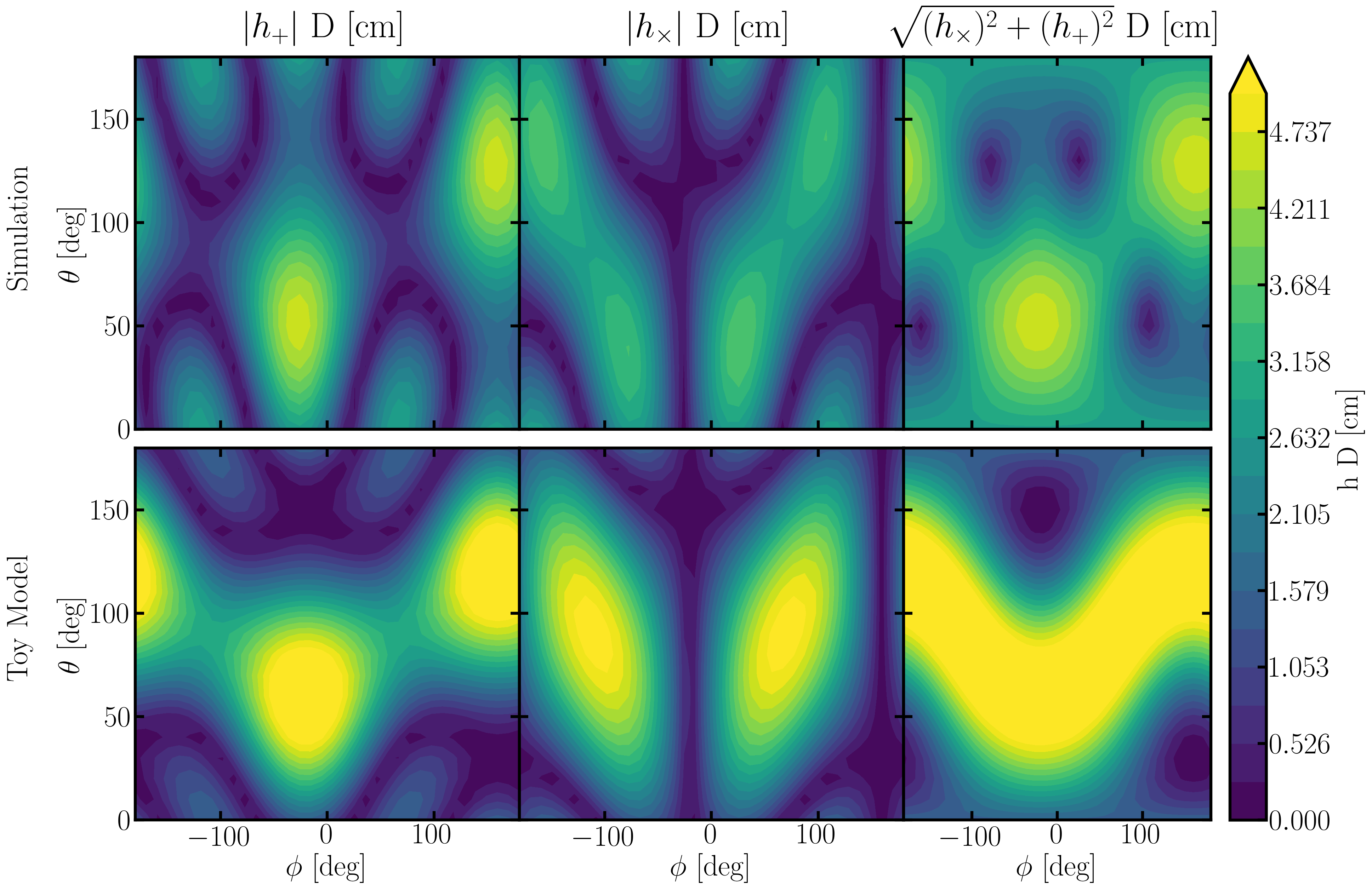}
    \caption{The angular dependency of the absolute value of the total GW signal from the simulation at the end of the second simulation phase, (tow row) compared with the angular distribution predicted by our prolate model (bottom row). The left column shows the plus polarization mode, the middle column shows the cross polarization mode, and the right column shows the gauge invariant root sum square of the two modes. The input parameters for the analytical model are those determined by our $\chi^2$-fit and are given in table~\ref{tab:Least Square}.}
    \label{fig:Sph Comp}
\end{figure*}

Table \ref{tab:Least Square} shows the minimized $\chi_{\nu}^{2}$ values and their associated parameters for the (gauge independent) memory produced ($\sqrt{h_{\times}^{2} + h_{+}^{2}}$) by the flow. We only fit the analytical models to the GW signal generated by the flow because the models are meant to represent the material ejected in the supernova and can, therefore, only capture the emission caused by asymmetric mass ejection. We leave it to further work, based on analysis of a larger set of numerical simulations, to develop more detailed models that can capture the neutrino generated emission as well. Connecting the properties of the ejected material to the memory it produces could be used to estimate the portion of the GW signal produced by asymmetric neutrino emission, which can be used to constrain the angular variability of the neutrino emission. Thus, by combining electromagnetic, GW, and neutrino observations more information can be extracted about the properties of a supernova explosion than what can be gleaned by any single channel alone. In order to quantify how well the toy models match the data, We provide a quantitative comparison to the values of the $\chi_{\nu}^{2}$ for the toy models in table \ref{tab:Least Square} with a $\chi_{\nu}^{2}$ estimation where there is no correlation at all: a Monte Carlo simulation where we compare the simulation to a sheet of random noise with expectation equal to the average of the memory values for the total gauge independent signal ($\mu = 12.64$) and standard deviation equal to the standard deviation of the same ($\sigma = 5.48$), and we find that $\langle\chi_{\nu}^{2}(\delta = 0)\rangle = 9.60$ with $\sigma = 1.01$ and $\langle\chi_{\nu}^{2}(\delta = 0.1)\rangle = 9.70$ with $\sigma = 1.04$. In \fig{fig:Sph Comp} we compare the angular dependence of the memory amplitude of W15-2 to the angular emission pattern of our prolate toy model with the input parameters of the model chosen per the $\chi_{\nu}^{2}$ analysis. The emission of the simulation appears to be fairly well represented by the analytical model, but the simulation shows a higher degree of variance between extrema than our analytical model. The analytical model roughly reproduces the location of emission local maxima and minima, but we are unable to capture the fine structure of the emission. This might be due to the symmetry constrains we impose on the models and the, effectively, low $\ell$-number nature of the models.

It is interesting that all the toy models give the same estimation of the axis breaking the explosion spherical symmetry. Estimating this axis is an interesting goal because there are indications that the early-stage explosion asymmetries are reflected in the late-stage dynamics \citep{2013A&A...552A.126W,2015A&A...577A..48W}. This goal has also already been investigated in the electromagnetic channel with Doppler shift measurements (see for example \citep{2016ApJ...833..147L}), which are mostly visible if the observer is perpendicular to the symmetry axis (in our toy models the simulations the GW amplitude appears on average larger for observers perpendicular to the explosion axis as well). In \citep{2016ApJ...833..147L} they observationally investigate the geometry of SN1987a 10000 days after the explosion and deduce the geometry of the explosion as a ``broken dipole", which resembles our prolate explosion model.

While in general the relationship between the neutrino, GW, and electromagnetic emission could be complicated, for rapidly rotating progenitors some degree of symmetry about the rotation axis is enforced which could simplify the connection between the three. We leave a systematic discussion of multi-messenger strategies for memory parameter estimation, including testing the validity of the toy models introduced here on other simulations, to future works.

\begin{table}
  \caption{The table shows the minimized least square values for the different models and their corresponding parameters.}
  \label{tab:Least Square}
  \begin{ruledtabular}
    \begin{tabular}{clll}
      \textbf{Model} & \textbf{Prolate} & \textbf{Oblate} & \textbf{Spheroid} \\
      \hline
      \multirow{4}{*}{Flow} & $M \times \alpha = 3$ & $M \times \alpha = 6$ & $M \times a = 2.3$,  $M \times c = 0.6$ \\
      & $\kappa = -20$ & $\kappa = -20$ & $\kappa = -180$ \\
      & $\mu = 30$ & $\mu = 30$ & $\mu = 90$ \\
      & $\chi^{2}_{r} = 0.64$ & $\chi^{2}_{r} = 0.64$ & $\chi^{2}_{r} = 1.15$ \\
    \end{tabular}
  \end{ruledtabular}
\end{table}

\subsection{Polarization in the Low Frequency Limit} \label{sec:low}

\begin{figure}
    \centering \includegraphics[width = 3.2 in]{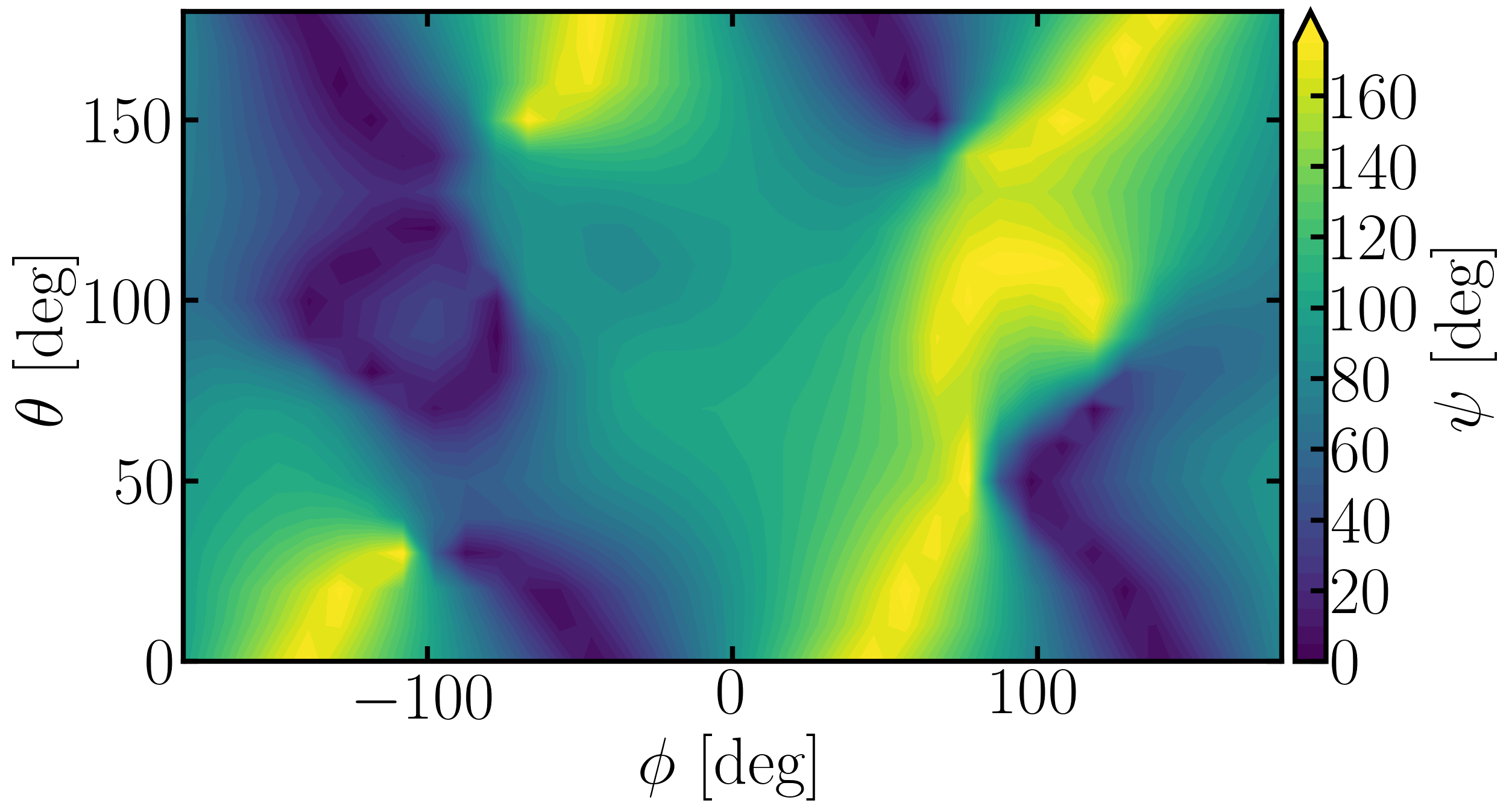}
    \caption{The rotation angle $\psi$ needed to transform to a coordinate system in which one of the polarization modes are zero, as a function of observer direction.}
    \label{fig:Polarization Total}
\end{figure}

\begin{figure}
    \centering \includegraphics[width=3.2 in]{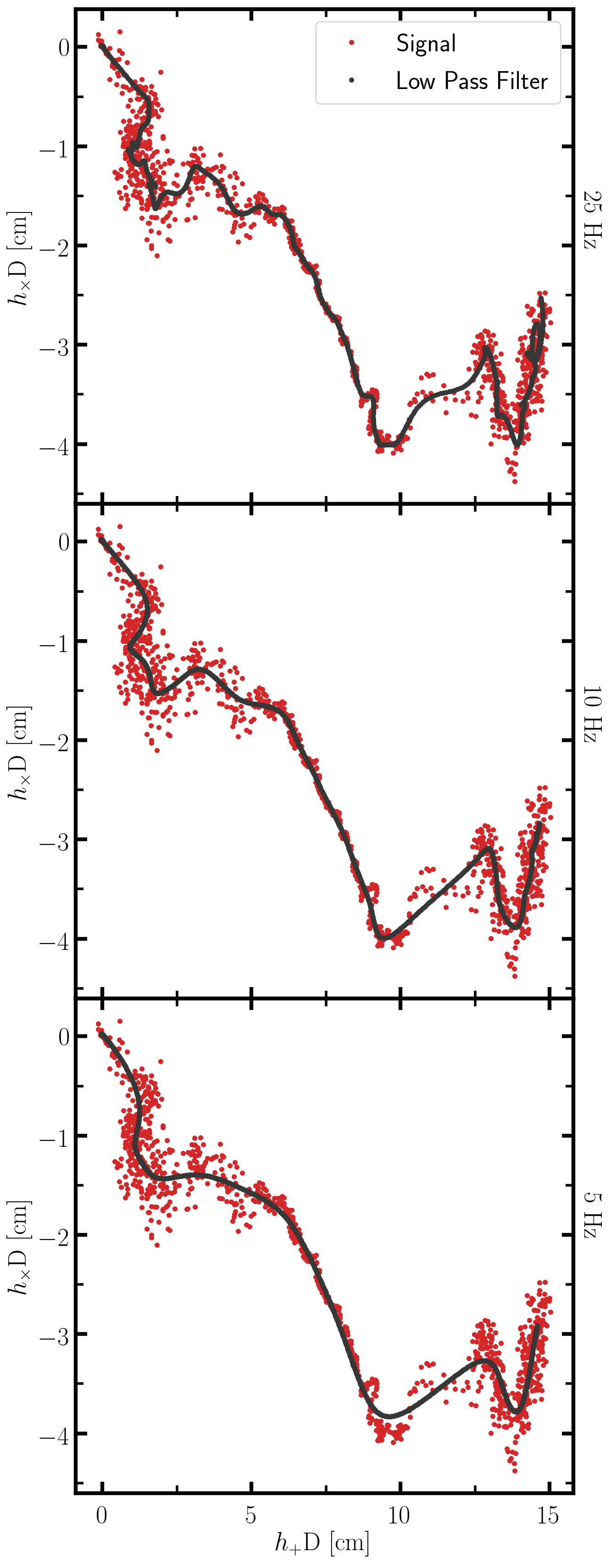}
    \caption{Pologram of a fixed observer direction ($\theta = 0 $, $\phi = -180$). Note that as the higher frequencies are cut off, the structure of the pologram does not change much. The major feature of the pologram is the approximately linear memory because its general diagonal shape.}
    \label{fig:Pologram}
\end{figure}

At any given time, the strain tensor can be written as a diagonal matrix by rotating the coordinate system by an angle $\psi$ around the GW traveling direction. In general, this angle will change over time. If the angle $\psi$ is not time-dependent, then the signal is labelled linearly polarized. In the low-frequency band and for the constant memory, the rotation angle will change more slowly, in comparison with the overall timescales of the GW burst. Consequently, it is expected that the low-frequency emission presented in this work asymptotically become linearly polarized. The two polarization modes of a linearly polarized signal in an arbitrary frame (defined by $\Phi$) obey the relation $h_{\times} = \alpha h_{+}$, where $\alpha$ is some arbitrary constant. For a linearly polarized signal, a scatter plot of $h_{\times}$ versus $h_{+}$ would produce a straight line following the diagonal of the figure (such scatter plots are known as ``polograms") \citep{2008CQGra..25k4029K}. In \fig{fig:Pologram}, we show a pologram for our signal, at each time-step, $t_i$ we plot $h_{\times}(t_{i})$ versus $h_{+}(t_{i})$. We applied a digital low-pass filter, cutting off all emission above a certain frequency limit, to the data before producing the plot. The three panels in \fig{fig:Pologram} represents three different cut-off frequencies: 5 Hz (bottom panel), 10 Hz (middle panel), and 25 Hz (top panel). The points lie closer and closer to the diagonal line as the cut-off frequency is reduced. In particular, the upper left and lower right portions of the plot smooth out. The panels in \fig{fig:Pologram} are produced from the signal one particular observer, albeit an arbitrary one, would observe and the rotation angle required to diagonalize the GW tensor will in general not be observer independent. \fig{fig:Polarization Total} shows the outgoing direction dependence of the rotation angle $\psi$ for the signal at a time equal to 1.3 s (at the end of simulation phase 1).

\section{Extending the Neutrino Generated Signal} \label{sec:Extension}
Since the neutrino transport is switched off at approximately 1.3 s post bounce, our GW signal is incomplete. At late times only the signal generated by mass motions can be calculated directly from the simulation. The GW signal associated with asymmetric neutrino emission must be extended by hand. The true properties of the neutrino emission, and consequently the GW emission, can only be determined by numerical simulations. To construct our extension function, we start with the expression for the GWs associated with anisotropic neutrino emission:

\begin{align}
    \label{eq:gw_nue}
    h_{\times/+}(t) = \frac{2G}{c^4 R} \int_{0}^{t}\alpha_{\times/+}(\tau, \beta, \gamma) L_{\nu} (\tau) \mathrm{d}\tau,
\end{align}

where $L_{\nu}$ is the total neutrino luminosity of the star at time $\tau$, and $\alpha_{\times/+}(\tau, \beta, \gamma)$ are anisotropy parameters which give a qualitative measure of the anisotropy of the neutrino emission (see \cite{2012A&A...537A..63M} for a detailed derivation of these quantities \footnote{Note that \cite{2012A&A...537A..63M} use $\Lambda$ in place of $L_{\nu}$}). The observer's orientation to the source is determined by the two angles $\beta$ and $\gamma$. From spherically symmetric simulations \citep{1999ApJ...513..780P, 2001ApJ...562..887T}, we know that the total neutrino luminosity during the cooling phase of the PNS is well approximated by an exponential function,
\begin{equation}
    \label{eq:lum1.3}
    L_{\nu}^{E}(t) = C t^{-n},
\end{equation}
with $n \sim 1$. The explosion is well underway at 1.3 s after bounce and the accretion rate has dropped significantly, and the luminosity should be fairly well approximated by the cooling luminosity. The constant $C$ can be determined by requiring that the analytical extension of the neutrino luminosity is equal to the neutrino luminosity extracted at the end of the simulation's first phase

\begin{equation}
    C = 1.3^n L_{\nu}(1.3).
\end{equation}

\begin{figure*}
    \centering \includegraphics[width=0.99\textwidth]{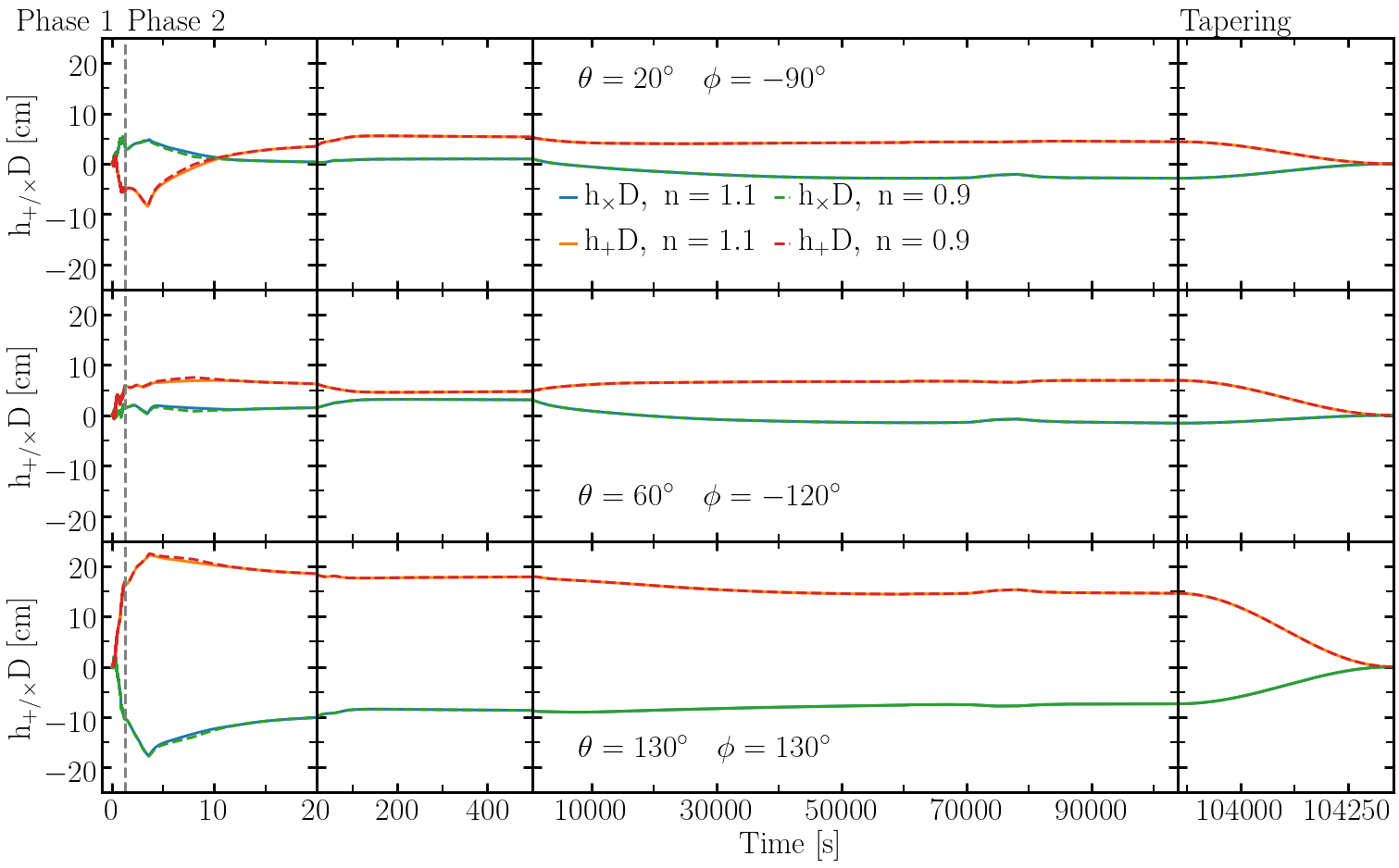}
    \caption{The total GW signal, the sum of the GWs generated by asymmetric neutrino emission and by matter motions, from both simulation phases. The neutrino signal in phase two has been extrapolated from the signal during phase one, as described in section~\ref{sec:Extension}. The observer directions are the same as in \fig{fig:amps} and \fig{fig:amps_late}. Each row represents one observer direction, which is denoted by the azimuthal ($\phi$) and polar ($\theta$) angle, in the spherical coordinate system of the hydrodynamic simulation.  Time is given in seconds after core bounce.}
    \label{fig:amps_ext}
\end{figure*}

Furthermore, the total energy radiated away from the PNS by neutrinos is $\sim 3 \times 10^{53}$ erg \citep{1990RvMP...62..801B}. The choice of $3 \times 10^{53}$ ergs is somewhat arbitrary and in reality, the exact amount of energy radiated by neutrinos will depend on the details of the core-collapse explosion. Different progenitors will radiate different amounts, but it is expected that the total energy lost by neutrinos should be around $10^{54}$ erg. The goal of our study is to investigate the overall signal properties, not the exact details of one specific progenitor’s core-collapse and, therefore, the choice we make for the total radiated energy must only be in the right ballpark. With a total energy loss by neutrinos of $\sim 3 \times 10^{53}$ erg, we have
\begin{equation}
    \label{eq:toterg}
    \int_{0}^{1.3} L_{\nu}(\tau) \mathrm{d}\tau + \int_{1.3}^{t_f} L_{\nu}^{E}(\tau) \mathrm{d}\tau \sim 3 \times 10^{53} \ \mathrm{erg}.
\end{equation}

The end time of the neutrino burst ($t_{f}$) can be found by combining \eqref{eq:lum1.3} and \eqref{eq:toterg}.

\begin{equation}
    t_f = \sqrt[1-n]{ \Big(3\times 10^{53} - L_{\nu}^N\Big)\frac{1-n}{C} + 1.3^{1-n}},
\end{equation}
where we have introduced the quantity $L_{\nu}^N = \int_{0}^{1.3} L_{\nu}(\tau) \mathrm{d}\tau \approx 8.366 \times 10^{52} \ \mathrm{erg} $ to simplify the expression. The exact value of $t_f$ depends on $n$, but for reasonable $n$-values we find that $t_f \sim 10$ s. If $n$ is equal to 1.1 then $C \approx 1.12 \times 10^{53} \mathrm{ erg \ s^{1.1}}$ and $t_f \approx 11.95 \ s$. When $n$ is 0.9 we find that $C \approx 1.06 \times 10^{53} \mathrm{ erg \ s^{0.9}}$ and $t_f \approx 7.98s$.

While the exact details depend on the observer orientation relative to the source, large changes in the amplitude generated by the matter signal take place between 1.3 and 20 s. The absolute value of the amplitude can change by more than a factor of two. Furthermore, the signal continues to evolve as the shock front propagates through the progenitor (see \fig{fig:amps_late}). In other words, the matter generated signal evolves on time scales far longer than those of the neutrino generated GW emission. The time evolution of the anisotropy parameters of the neutrino luminosity, and thus the GW signal from the neutrinos, are uncertain. However, there are two general cases for the GW production: the GW signal gains energy from the neutrino emission, or the anisotropy parameters evolve in such a way that the neutrino generated GW signal tapers off to zero. 
Since we are interested in studying the case where a memory signal arises and since we recognize that the anisotropy parameters can only be determined from numerical simulations, we take a pragmatic approach to extending the GW signal. We assume that the anisotropy parameters remain constant after 1.3 s. In this case, we get the following signal extension 
\begin{align}
    \label{eq:gw_ex1}
    h_{\times/+}^e(t_e) &= h_{\times/+}^{1.3} + \frac{2G}{c^4 R} \alpha_{\times/+}^c\int_{1.3}^{t_e} C \tau^{-n} (\tau) \mathrm{d}\tau \nonumber \\ 
    & = h_{\times/+}^{1.3} + \frac{2G}{c^4 R} \frac{\alpha_{\times/+}^c C}{1-n} \bigg(t_e^{1-n}-1.3^{1-n} \bigg).
\end{align}

Here $h_{\times/+}^{1.3}$ denotes the GW signal at the time step before neutrino transport was switched off, $\alpha_{\times/+}^c$ represents the constant anisotropy parameters, and $t_e$ is used to represent times after the neutrino transport was switched off. Before 1.3 s, the absolute value of the anisotropy parameters is typically $\sim 10^{-5} - 10^{-3} $ \citep{2012A&A...537A..63M}, we set $\alpha_{\times/+}^c = h_{\times/+}^{1.3}/ |h_{\times/+}^{1.3}| \times 10^{-4}$.

In \fig{fig:amps_ext} we show the total GW signal, generated by the matter and by the emission of neutrinos, as a function of time for the three same observer directions as shown in \fig{fig:amps}. The signal is shown for two different values of $n$. Firstly, we see that changing $n$ does not lead to notable differences in the signals. Secondly, comparing \fig{fig:amps_late} and \fig{fig:amps_ext} demonstrates that the signal generated from the flow alone can differ significantly from the signal when asymmetric neutrino emission is taken into account. Including the neutrino generated signal can switch the sign of the two polarization modes (this is easily seen by comparing the middle rows of \fig{fig:amps_late} and \fig{fig:amps_ext}). Generally, we see the same trend that we saw for the emission generated by the flow alone, the signal reaches a maximum or minimum during the first 500 s of the second simulation phase and then tends towards a specific value as the simulation progresses. When the neutrino generated signal is included, the final value that the signal tends towards is non-zero. The effect of taking neutrino generated emission mostly changes the amplitude of the signal, which is what we expect from our simple extension method. Note, the last column in \ref{fig:amps_ext} shows the tapering of the signal with a tapering frequency of $1$ mHz (see section \ref{sec:Tapering}).

\section{Tapering} \label{sec:Tapering}
If the amplitude of a GW signal is non-zero at the end of a numerical simulation, either for physical or numerical reasons, the abrupt jump from some finite value to zero will induce artefacts in the energy spectrum of the signal. Such issues could be handled by applying a windowing function to the signal with precise understanding of the impact at different frequencies. However, such methods are not applicable on the signal itself since they remove or distort energy content in the sensitive band of the instrument. We, therefore, extend our signals with a function that tapers to zero over some finite time scale.

\begin{figure}
    \centering \includegraphics[width = 3.2 in]{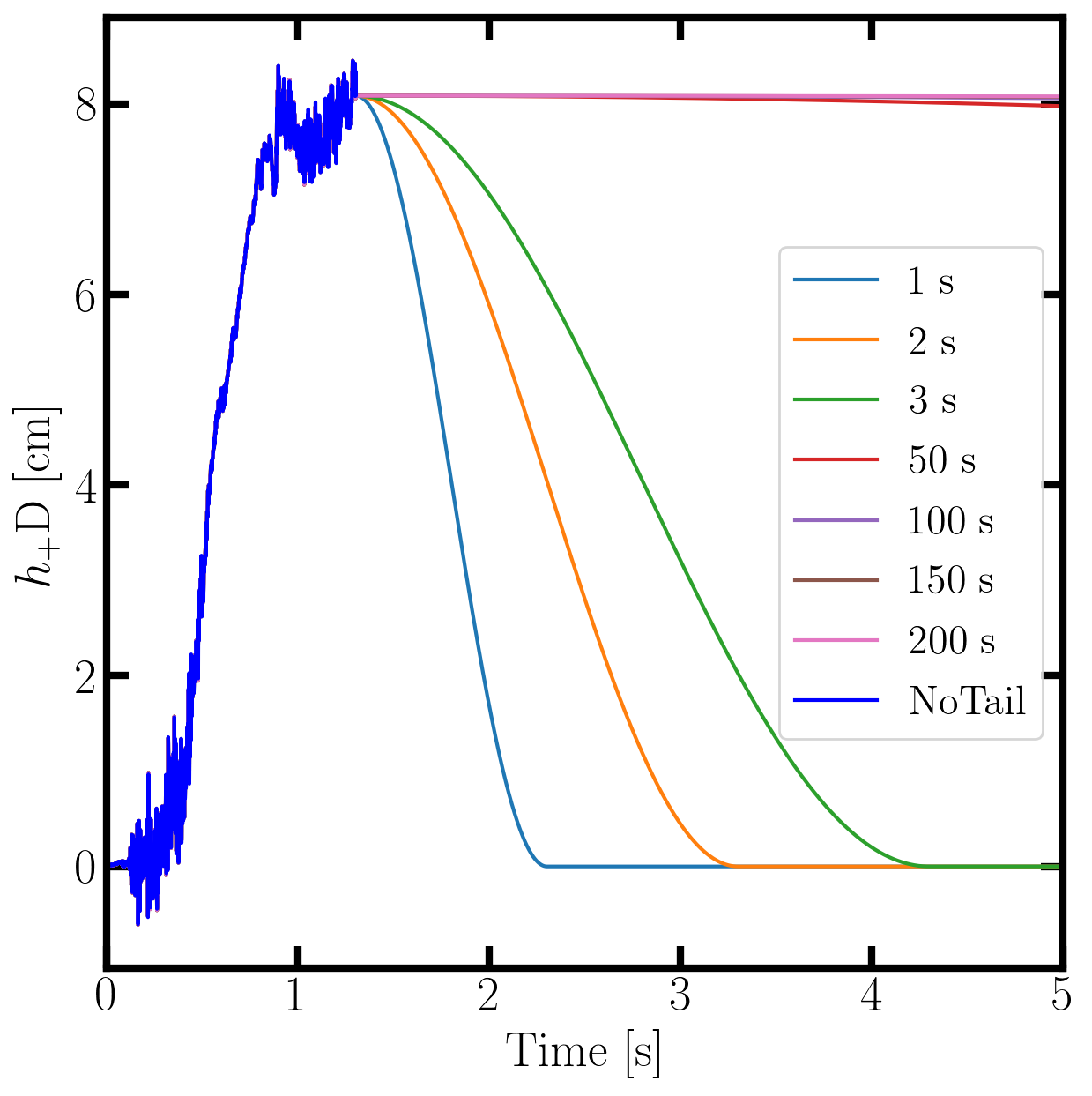}
    \caption{Demonstration of various taperings where we vary the time scale over which the tapering function settles down to zero. Note that the curves are difficult to tell apart, in this figure, for taperings longer than 3 s. The darker blue curve shows the GW signal for a random observer up to 1.3 s.}
    \label{fig:Tails Time}
\end{figure}

We use the following tapering function
\begin{equation}
    \label{eq:taperf}
    h^{\text{tail}}_{\times/+} = \frac{h^{\text{end}}_{\times/+}}{2}\big[ 1 + \cos(2\pi  f_t(t-t^{\text{end}}))\big],
\end{equation}
where $t^{\text{end}}$ is the duration of the simulation, $h^{\text{end}}$ represents the signal value at $t^{\text{end}}$, and $f_t$ is the frequency of the tapering function. The subscript ${\times/+}$ denotes the cross and plus polarization modes, respectively. In \fig{fig:Tails Time} we show the signal observed by a specific observer $(0^{o}, -180^{o})$ during the first simulation phase (the dark blue curve) extended by different tapering functions. The tapering functions only differ in the choice of $f_t$, which was step-wise decreased from $1$ Hz to $1/200$ Hz (\eq{eq:taperf}).

\begin{figure*}
    \centering \includegraphics[width=0.99\textwidth ]{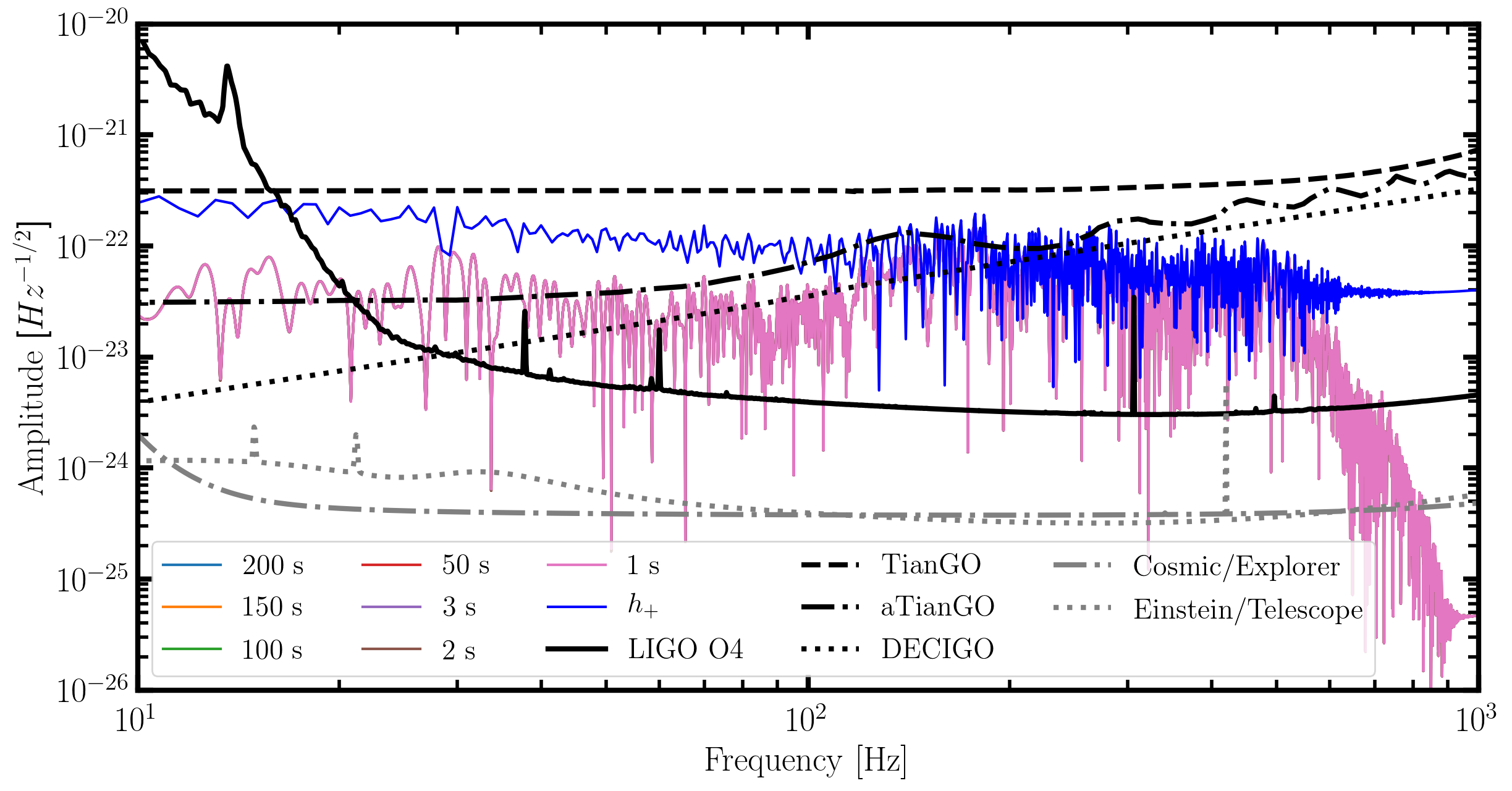}
    \caption{Blue line: the spectral energy density of the plus polarization mode of the GW emission emitted in the direction defined by $(\theta,\phi) = (0^{o}, -180^{o})$ without tapering. The black and gray curves show the sensitivity curves of various detectors (as indicated by the label). The source is assumed to be at a distance of 1 kpc. The curves labeled by a specific amount of seconds show the spectral energy density of the signal after a tapering of the specified time duration has been added. This figure shows the curves between 10 and 1000 Hz. Notice that for frequencies $>10$ Hz, the signals with the addition of the tails are indistinguishable.}
    \label{fig:No_LISA_1kpc_SNR}
\end{figure*}

\begin{figure*}
    \centering \includegraphics[width=0.99\textwidth ]{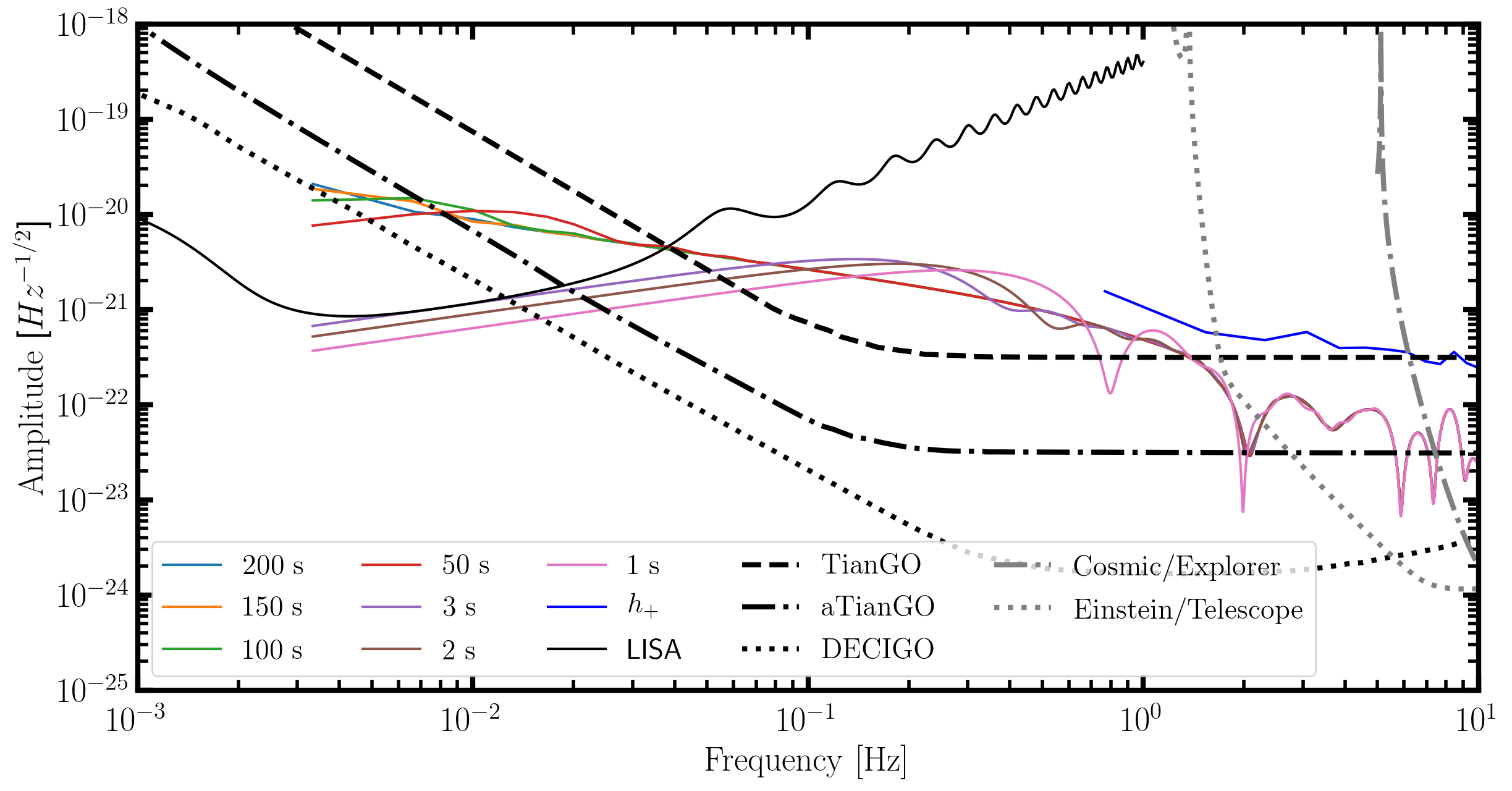}
    \caption{Blue line: the spectral energy density of the plus polarization mode of the GW emission emitted in the direction defined by $(\theta,\phi) = (0^{o}, -180^{o})$ without tapering. The black and gray curves show the sensitivity curves of various detectors (as indicated by the label). The source is assumed to be at a distance of 1 kpc. The curves labeled by a specific amount of seconds show the spectral energy density of the signal after a tapering of the specified time duration has been added.
      This figure shows the curves below 10 Hz.}
    \label{fig:LISA_1kpc_SNR}
\end{figure*}

In \fig{fig:No_LISA_1kpc_SNR} we quantify the effect of tapering the GW signals. We show the Fourier transform of the signal from the first phase of the simulation without (dark blue curve) and with tapering compared with the projected noise curves of LIGO O4, TianGO, aTianGO and DECIGO, assuming that the source is at a distance of 1 kpc. The un-tapered signal (dark blue curve) has a nonphysical excess of energy above $\sim 700$ Hz, which disappears once tapering is introduced. Failure to properly taper the signal can lead to an overestimation of the SNR for any given model. In \fig{fig:LISA_1kpc_SNR}, the frequency range has been restricted to 1 mHz to 10 Hz and we show the noise curves of LISA, TianGO, aTitanGO, and DECIGO. The source is assumed to be at a distance of 1 kpc.

It is also important to note that any tapering function will add energy in a frequency band centered around $f = \frac{1}{t}$ where $t$ is the time scale of the tapering duration. This means that if we do not want to add nonphysical energy around a frequency $f$ we need to be confident that the tapering is reliable over a duration $t=\frac{1}{f}$. In rough terms as long as the tapering is physically realistic over time scales a bit longer that the inverse of the frequency where the spectrum of the signal goes below the spectral amplitude of the noise of a specific interferometer of interest, we can trust the detectable portion of the signal. For space-based detectors
one has to be careful when choosing the length of the tapering to avoid
injecting nonphysical energy within the sensitivity range of the detectors. This is semi-quantitative because the exact criteria depends on the algorithm adopted for the memory extraction. This time scale will also depend on the exact amplitude of the memory production from a specific observer orientation. While we do not describe a detailed landscape of memory amplitudes and SNRs, we would like to point out that this model is fairly conservative, for example in \citep{2020MNRAS.492.4613O} the amplitude is an order of magnitude larger. It is also worth pointing out that the frequency where the intersection occurs will change, with respect to the distance of the source, depending on the amplitude of the memory and the slope of the noise floor.

The SNR for three future detectors (aLIGO O4, Lisa, and TianGO) are presented in table \ref{tab:SNR}. Notice that the SNR for LIGO is overestimated without tapering and therefore it may produce unrealistic predictions. Furthermore, we
estimate the SNR using the signal from the first simulation phase,
which means that the SNR is determined by the properties of the tail
(see \fig{fig:Tails Time}). Without a proper extension of the signal
or a simulation covering the full duration of the core-collapse
event, it is difficult to accurately estimate the SNR for spaced-based
detectors.

\begin{table}
\caption{\label{tab:SNR}The Signal to Noise Ratio of a randomly oriented GW signal scaled to 1 kpc is shown. Note the reliance of the SNR on the addition of $0 s$, $1 s$, and $100 s$ tails. For larger distances, the SNR is scaled by a factor of $\frac{1}{D}$, for $D$ in kpc.}
\begin{ruledtabular}
\begin{tabular}{cll}
 \textbf{Detector} & \textbf{Tail} & \textbf{1 kpc} \\ 
 \hline
 \multirow{3}{*}{aLIGO O4} & 0 s & 15.851 \\ & 1 s & 4.446 \\ & 100 s & 4.462 \\ 
 
 \hline
 
 \multirow{3}{*}{Cosmic Explorer} & 0 s & 495.8 \\ & 1 s & 120.4\\ & 100 s & 120.4  \\ 
   
 \hline
 
 \multirow{3}{*}{Einstein Telescope} & 0 s & 312.1 \\ & 1 s & 65.4  \\ & 100 s & 65.4  \\ 
  
 \hline
 
 \multirow{3}{*}{TianGO} & 0 s & 5.445  \\ & 1 s & 8.746 \\ & 100 s & 6.409  \\ 
  
 \hline
 
  \multirow{3}{*}{DECIGO} & 0 s & 961.8  \\ & 1 s & 977.7 \\ & 100 s & 623.6  \\ 
  
 \hline
 
 \multirow{3}{*}{LISA} & 0 s & $10^{-3}$ \\ & 1 s & 0.833  \\ & 100 s & 20.171 \\
\end{tabular}
\end{ruledtabular}
\end{table}

\section{Conclusion} \label{sec:Conclusion}
The linear memory of gravitational waves from core-collapse supernovae will be detectable utilizing future low-frequency interferometers. In this study, we recommend the use of a combination of tapering and extensions on GW signals from core-collapse supernova simulations. This will reduce the artificial injection of energy at high frequencies induced by the truncation of the simulation. In this paper we discussed how this low-frequency portion of the signal is expected to be predominantly linearly polarized. This means that tuning of GW burst detection algorithms like cWB favoring linear signals might be appropriate \citep{2008CQGra..25k4029K}.

It would, at this point, be prudent to once more point out that our analysis is based on a numerical simulation which utilizes a fairly approximate neutrino transport scheme and replaces the inner PNS with a time evolving boundary condition. The simulation was tuned to produce an explosion and neutrino transport was shut off in the second simulation phase. The approximate nature of the neutrino transport likely produces local artefacts, but we are mainly with the global and large-scale properties of the simulations. Local variations are averaged out when performing the integrals necessary to calculate the GW and neutrino signals. The signal generate by asymmetric neutrino emission should be most adversely affected by the approximations of the simulation, while the low-frequency matter generated signal should be accurate. Furthermore, the main goal of our work is to study a signal class and not an individual signal. We do not require an exact signal, but rather information about the overall morphology of these types of signals. However, it is important to note that the signal presented here originates from a simulation with approximate neutrino transport and will, therefore, will contain some inaccuracies. We refer the interested reader to the discussion in \cite{2012A&A...537A..63M}, in particular the two last paragraphs of the conclusion.

\fig{fig:amps} demonstrates that both the sign and the amplitude depend on from which direction the source is viewed. However, the amplitude, as seen in \fig{fig:angular_dis}, is used for interpretations of the angular dependence since it is gauge invariant, but not observer invariant. The study of the angular dependence of the amplitude of the memory indicates that to perform injections with random orientations and present average results a coarser angular resolution than $\sim10$ degrees could sample the variability of the memory. The angular dependence of the matter contribution to the GW production is more regular than the neutrinos are. In this paper, we do not investigate different waveforms with different memory amplitudes but leave this to a future study. However, we notice waveforms with up to 10 times the amplitude shown in this paper (for example see \citep{2020MNRAS.492.4613O}).

The study of different possible taperings shows that as long as the tapering is longer than 10 seconds the in-band spectrum for LIGO will not be affected. However, for future generation detectors, like LISA, the tapering needs to be physically correct up to even longer time scales.
The dominant contribution to the steep climb of the noise at low frequency, is the response to linear and non-linear vibrations, mainly from the ground. This is a limitation that can be improved with further insulation systems (technical noise) and not a fundamental physics limitation. Furthermore, because the detectors' response to noise sources are both linear and non-linear, there are recent indications that Volterra expansions and non-linear couplings could be used to reduce the noise in this frequency range by removing parts of the noise predictable from auxiliary monitoring sensors (see for example \cite{2020PhRvD.101d2003V}). The conclusions of this work for ground-based interferometers are somehow conservative for what could be achieved.  One aspect that we do not explore here is how to manage the motion of the Earth for longer signal durations. We leave a detailed discussion of this topic for future work.

\section{Acknowledgements} \label{sec:Ack}

This research has made use of data, software and/or web tools obtained from the Gravitational Wave Open Science Center, a service of LIGO Laboratory, the LIGO Scientific Collaboration and the Virgo Collaboration. Specifically, we acknowledge Martina Muratore for her assistance in understanding and implementing the LISA noise floor and we acknowledge conversations and assistance from Cecilia Lunardini regarding the extensions for the neutrino generated GWs. MZ was supported by NSF Grant No.PHY-1806885. We gratefully acknowledge the support of LIGO and Virgo for the provision of computational resources.  Plots were produced using \citep{2007CSE.....9...90H} and computations were preformed with \cite{2020Natur.585..357H}.

\newpage

\appendix

\section{Directionality Derivations}

\subsection{Prolate explosion with arbitrary directionality}

Let $\hat{n} = (\cos{\phi} \sin{\theta}, \sin{\phi} \sin{\theta}, \cos{\theta})$ be the arbitrary direction of propagation of the GW, the density can be written as

\begin{equation}
\begin{split}
\rho & =
\frac{M}{2}(\delta(x-f(t)\hat{n}\cdot{}\hat{i})\delta(y-f(t)\hat{n}\cdot{}\hat{j})\delta(z-f(t)\hat{n}\cdot{}\hat{k}))
\\ & + \frac{M}{2}(- \xrightarrow{} +) \\
\end{split}
\end{equation}

We solve for the individual moments of inertia

\begin{equation}
\begin{split}
I_{11} & = Mf(t)^{2}\cos^{2}{\phi} \sin^{2}{\theta}
\end{split}
\end{equation}

\begin{equation}
I_{22} = Mf(t)^{2}\sin^{2}{\phi} \sin^{2}{\theta}
\end{equation}

\begin{equation}
I_{33} = Mf(t)^{2}\cos^{2}{\theta}
\end{equation}

\begin{equation}
\begin{split}
I_{12} & = I_{21} = M f(t)^{2} \cos{\phi} \sin{\phi}\sin^{2}{\theta}
\end{split}
\end{equation}

\begin{equation}
I_{13} = I_{31} = M f(t)^{2} \cos{\phi} \sin{\phi} \cos{\theta}
\end{equation}

\begin{equation}
I_{23} = I_{32} = M f(t)^{2} \sin{\phi} \sin{\theta}\cos{\theta}
\end{equation}
or collectively

\begin{widetext}
\begin{equation}
I = M f(t)^{2}
\begin{pmatrix}
\cos^{2}{\phi} \sin^{2}{\theta} & \cos{\phi} \sin{\phi} \sin^{2}{\theta} & \cos{\phi} \sin{\phi}
\cos{\theta} \\ \cos{\phi} \sin{\phi} \sin^{2}{\theta} & \sin^{2}{\phi} \sin^{2}{\theta} &
\sin{\phi} \cos{\theta} \sin{\theta} \\ \cos{\phi} \sin{\phi} \cos{\theta} & \sin{\phi} \cos{\theta}
\sin{\theta} & \cos^{2}{\theta} \\
\end{pmatrix}
\end{equation}
\end{widetext}

\subsection{Oblate explosion with arbitrary directionality} \label{sec:Oblate Inertia}

We consider the scenario where the explosion propagates radially in a plane perpendicular to the axis $\hat{n}$. Therefore, the locations where $\rho \neq 0$ have to verify $\vec{x} \cdot
\hat{n} = 0$, and $\rho \propto \delta (\vec{x} \cdot \hat{n})$.
 Note that the integral of the density overall space returns the mass.

\begin{equation}
\rho = \frac{M}{2 \pi r} \delta (r - f(t)) \delta (\vec{x} \cdot \hat{n})
\end{equation}

Note that $|\vec{\nabla} (\vec{x} \cdot \hat{n})| = |\hat{n}| = 1$, so we now choose $d \sigma (\vec{x})$ as the 2-D representation $r dr d\phi$ in the plane. We define the density $\rho$ below. We define a rotation matrix $R$ that transforms a vector from the Cartesian reference
frame to the cylindrical frame where $\hat{n}$ is the z-axis, such that

\begin{equation}
\vec{x} \cdot \hat{i} = (\vec{x} R^{-1}) (R \hat{i}) = \sum_{l = 1}^{3} (\vec{x} R^{-1})_{l}
(R\hat{i})_{l}
\end{equation}

The moment of inertia is

\begin{equation}
\begin{split}
I_{ij} & = \int (\vec{x} R^{-1})_{l} (R \hat{e}_{i})_{l} (\vec{x} R^{-1})_{m} (R \hat{e}_{j})_{m}
\rho d^{3}x \\ & = (R \hat{e}_{i})_{l} (R \hat{e}_{j})_{m} \Tilde{I}_{lk}
\end{split}
\end{equation}

Defining $R$ explicitly in terms of $\hat{i}$, $\hat{j}$, and $\hat{k}$ and $\hat{n}$, $\hat{\xi}$,
and $\hat{\eta}$ we get

\begin{equation}
R =
\begin{pmatrix}
\hat{n} \cdot \hat{i} & \hat{n} \cdot \hat{j} & \hat{n} \cdot \hat{k} \\ \hat{\xi} \cdot \hat{i} &
\hat{\xi} \cdot \hat{j} & \hat{\xi} \cdot \hat{k} \\ \hat{\eta} \cdot \hat{i} & \hat{\eta} \cdot
\hat{j} & \hat{\eta} \cdot \hat{k} \\
\end{pmatrix}
\end{equation}
with $R^{T} R = R R^{T}, R^{T} = R^{-1}$.

We now investigate the moment of inertia again. Where $\vec{y} = R\vec{x}$. Explicitly $I_{11}$
becomes:

\begin{equation}
\begin{split}
I_{11} = (R \hat{i})_{l} (R \hat{i})_{m} \int y_{l} y_{m} \rho d^{3}y = (R \hat{i})_{l} (R
\hat{i})_{m} \Tilde{I}_{lk}
\end{split}
\end{equation}

$\Tilde{I}_{lk}$ is invariant from the rotation matrix and is defined by:

\begin{equation}
\Tilde{I}=
\begin{pmatrix}
0 & 0 & 0 \\ 0 & \Tilde{I}_{22} & \Tilde{I}_{23} \\ 0 & \Tilde{I}_{32} & \Tilde{I}_{33} \\
\end{pmatrix}
\end{equation}

Note that any element with $l,k = 1$ is a zero, because $\delta(\hat{n} \cdot \vec{x}) =
\delta(y_{1})$. Also, due to the symmetry of the distribution $\Tilde{I}_{23} = \Tilde{I}_{32}$ and
$\Tilde{I}_{22} = \Tilde{I}_{33}$. We calculate $\Tilde{I}_{23}$ and $\Tilde{I}_{22}$ below.

\begin{equation}
\begin{split}
\Tilde{I}_{22} =\frac{M}{2} f^{2}(t) \\
\end{split}
\end{equation}

\begin{equation}
\begin{split}
\Tilde{I}_{23} = \Tilde{I}_{32} = 0 \\
\end{split}
\end{equation}

\begin{equation}
\Tilde{I} =
\begin{pmatrix}
0 & 0 & 0 \\ 0 & \frac{M}{2} f^{2}(t) & 0 \\ 0 & 0 & \frac{M}{2} f^{2}(t) \\
\end{pmatrix}
\end{equation}

We can now compute $I_{\alpha \beta}$, knowing it is symmetric we calculate $I_{11}$, $I_{22}$,
$I_{33}$, $I_{12}$, $I_{13}$, and $I_{23}$. Beginning with the diagonal terms.

\begin{equation}
\begin{split}
I_{11} & = (\hat{n} \cdot \hat{i}, \hat{\xi} \cdot \hat{i}, \hat{\eta} \cdot \hat{i})
\begin{pmatrix}
0 & 0 & 0 \\ 0 & 1 & 0 \\ 0 & 0 & 1 \\
\end{pmatrix}
\begin{pmatrix}
\hat{n} \cdot \hat{i} \\ \hat{\xi} \cdot \hat{i} \\ \hat{\eta} \cdot \hat{i} \\
\end{pmatrix}
\frac{M}{2} f^{2}(t) \\ & = \frac{M}{2} f^{2}(t) [1 - \sin^{2}(\theta_{n}) \cos^{2}(\phi_{n})]
\end{split}
\end{equation}

\begin{equation}
\begin{split}
I_{22} = \frac{M}{2} f^{2}(t) [1 - \sin^{2}(\theta_{n}) \sin^{2}(\phi_{n})] \\
\end{split}
\end{equation}

\begin{equation}
\begin{split}
I_{33} = \frac{M}{2} f^{2}(t) \sin^{2}(\theta_{n})\\
\end{split}
\end{equation}

\begin{equation}
\begin{split}
I_{12} =\frac{M}{2} f^{2}(t) [(\hat{\xi} \cdot \hat{i})(\hat{\xi} \cdot \hat{j}) + (\hat{\eta} \cdot
  \hat{i})(\hat{\eta} \cdot \hat{j})]
\end{split}
\end{equation}

\begin{equation}
\begin{split}
I_{12} = -\frac{M}{2} f^{2}(t) \cos{(\phi_{n})} \sin{(\phi_{n})} \sin^{2}{(\theta_{n})}] \\
\end{split}
\end{equation}
\begin{equation}
\begin{split}
I_{13} = -\frac{M}{2} f^{2}(t) \sin{(\theta_{n})}\cos{(\theta_{n})}\cos{(\phi_{n})}] \\
\end{split}
\end{equation}
\begin{equation}
\begin{split}
I_{23} = -\frac{M}{2} f^{2}(t) \sin{(\theta_{n})}\cos{(\theta_{n})}\sin{(\phi_{n})} \\
\end{split}
\end{equation}

We now define the entire momenta tensor.

\begin{widetext}
\begin{equation}
I = \frac{M}{2} f^{2}(t)
\begin{pmatrix}
1 - \cos{\phi} \sin^{2}{\theta}& - \cos{\phi} \sin{\phi} \sin^{2}{\theta} & \cos{\phi} \cos{\theta}
\sin{\theta} \\ - \cos{\phi} \sin{\phi} \sin^{2}{\theta} & 1 - \sin^{2}{\phi} \sin^{2}{\theta} & -
\sin{\phi} \cos{\theta} \sin{\theta} \\ - \cos{\phi} \cos{\theta} \sin{\theta} & - \sin{\phi}
\cos{\theta} \sin{\theta} & \sin^{2}{\theta} \\
\end{pmatrix}
\end{equation}
\end{widetext}

\subsection{Prolate spheroid density about the z-axis}
\label{sec:ZSph}
The generic equation for a prolate spheroid requires a surface that expands symmetrically in two axes, and independently in the third. For this first example we will assume symmetry in the $x\text{-}y$ plane, and can therefore define the surface as:

\begin{equation}
\frac{x^{2} + y^{2}}{a^{2}} + \frac{z^{2}}{c^{2}} = f^{2}
\end{equation}

We define the density of the prolate spheroid surface symmetric about the z-axis as:

\begin{equation}
\rho = \frac{M}{2\pi a^{2}(t) c(t) f(t)} \delta \bigg(\frac{x^{2} + y^{2}}{a^{2}(t)} +
\frac{z^{2}}{c^{2}(t)} - f^{2}(t)\bigg)
\end{equation}

As before we calculate the moment of inertia:

\begin{equation}
\begin{split}
I_{11} =\frac{M}{3} a^{2}(t) f^{2}(t) \\
\end{split}
\end{equation}

\begin{equation}
\begin{split}
I_{22} = \frac{M}{3} a^{2}(t) f^{2}(t) \\
\end{split}
\end{equation}

\begin{equation}
\begin{split}
I_{33} = \frac{M}{3} c(t) f^{2}(t) \\
\end{split}
\end{equation}

\begin{equation}
\begin{split}
I_{12} = I_{13} = I_{23}= 0 \\
\end{split}
\end{equation}

Therefore, we can define the inertial tensor as

\begin{equation}
\label{eq:Inertial Tensor}
\begin{pmatrix}
\frac{M}{3} a^{2}(t) f^{2}(t) & 0 & 0 \\ 0 & \frac{M}{3} a^{2}(t) f^{2}(t) & 0 \\ 0 & 0 &
\frac{M}{3} c(t) f^{2}(t) \\
\end{pmatrix}
\end{equation}

This is equivalent to

\begin{equation}
I_{Spheroid} = \frac{2}{3} \bigg(a^{2}(t) I_{Oblate} + c^{2}(t) I_{Prolate} \bigg)
\end{equation}

\subsubsection{Arbitrary Explosion Angle}

\label{sec:ArbSph}

By following the same process as in \ref{sec:Oblate Inertia}, we find the inertial tensor of a
spheroidal explosion with an arbitrary explosion axis as

\begin{widetext}
\begin{equation}
\label{eq:Sph}
{\scriptsize {\tfrac{f^2 M}{3}}
\begin{pmatrix}
 2 \left(a^2 \cos ^2(\theta ) \cos ^2(\phi )+2 c^2 \sin ^2(\theta ) \cos ^2(\phi )+a^2 \sin ^2(\phi
 )\right) & - \left(a^2-2 c^2\right) \sin ^2(\theta ) \sin (2 \phi ) & - \left(a^2-2 c^2\right) \cos
 (\phi ) \sin (2 \theta ) \\ - \left(a^2-2 c^2\right) \sin ^2(\theta ) \sin (2 \phi ) & 2 \left(a^2
 \cos ^2(\phi )+\left(a^2 \cos ^2(\theta )+2 c^2 \sin ^2(\theta )\right) \sin ^2(\phi )\right) & -
 \left(a^2-2 c^2\right) \sin (2 \theta ) \sin (\phi ) \\ - \left(a^2-2 c^2\right) \cos (\phi ) \sin
 (2 \theta ) & - \left(a^2-2 c^2\right) \sin (2 \theta ) \sin (\phi ) & 2 \left(2 c^2 \cos ^2(\theta
 )+a^2 \sin ^2(\theta )\right)
\end{pmatrix}}
\end{equation}
\end{widetext}

\bibliography{Bib}


\end{document}